\documentclass[twocolumn]{aastex631-2}
\usepackage{amsmath}

\newcommand{\new}[1]{{\color{black} #1}}

\defcitealias{2021pdaa.book.....N}{NAP2023}

\begin{document}

\title{Multi-Amplifier Sensing Charge-coupled Devices for Next Generation Spectroscopy}

\author[0000-0001-8967-2281]{Kenneth W. Lin}
\affiliation{Department of Astronomy, University of California, Berkeley, CA 94720, USA}
\affiliation{Lawrence Berkeley National Laboratory, One Cyclotron Rd, Berkeley, CA 94720, USA}

\author[0000-0003-2285-1765]{Armin Karcher}
\affiliation{Lawrence Berkeley National Laboratory, One Cyclotron Rd, Berkeley, CA 94720, USA}

\author[0000-0001-9822-6793]{Julien Guy}
\affiliation{Lawrence Berkeley National Laboratory, One Cyclotron Rd, Berkeley, CA 94720, USA}

\author[0000-0001-7218-3457]{Stephen E. Holland}
\affiliation{Lawrence Berkeley National Laboratory, One Cyclotron Rd, Berkeley, CA 94720, USA}

\author{William F. Kolbe}
\affiliation{Lawrence Berkeley National Laboratory, One Cyclotron Rd, Berkeley, CA 94720, USA}

\author[0000-0002-3389-0586]{Peter E. Nugent}
\affiliation{Lawrence Berkeley National Laboratory, One Cyclotron Rd, Berkeley, CA 94720, USA}
\affiliation{Department of Astronomy, University of California, Berkeley, CA 94720, USA}

\author[0000-0001-8251-933X]{Alex Drlica-Wagner}
\affiliation{Fermi National Accelerator Laboratory, Batavia, IL 60510, USA}
\affiliation{Department of Astronomy \& Astrophysics, University of Chicago, Chicago, IL 60637, USA}
\affiliation{Kavli Institute for Cosmological Physics, University of Chicago, Chicago, IL 60637, USA}

\author[0000-0001-6396-2467]{Ana M. Botti}
\affiliation{Fermi National Accelerator Laboratory, Batavia, IL 60510, USA}
\affiliation{Kavli Institute for Cosmological Physics, University of Chicago, Chicago, IL 60637, USA}

\author{Javier Tiffenberg}
\affiliation{Fermi National Accelerator Laboratory, Batavia, IL 60510, USA}

\begin{abstract}
We present characterization results and performance of a prototype Multiple-Amplifier Sensing (MAS) silicon charge-coupled device (CCD) sensor with 16 channels potentially suitable for faint object astronomical spectroscopy and low-signal, photon-limited imaging. The MAS CCD is designed to reach sub-electron readout noise by repeatedly measuring charge through a line of amplifiers during the serial transfer shifts. Using synchronized readout electronics based on the DESI CCD controller, we report a read noise of 1.03 e- rms/pix at a speed of 26 $\mu$s/pix with a single-sample readout scheme where charge in a pixel is measured only once for each output stage. At these operating parameters, we find the amplifier-to-amplifier charge transfer efficiency (ACTE) to be $>0.9995$ at low counts for all amplifiers but one for which the ACTE is 0.997. This charge transfer efficiency falls above 50,000 electrons for the read-noise optimized voltage configuration we chose for the serial clocks and gates. The amplifier linearity across a broad dynamic range from $\sim$300--35,000 e- was also measured to be $\pm 2.5\%$. We describe key operating parameters to optimize on these characteristics and describe the specific applications for which the MAS CCD may be a suitable detector candidate.
\end{abstract}


\section{Introduction} \label{sec:intro}
Charge-coupled devices (CCDs) have played a pivotal role in scientific imaging applications including in dark matter direct detection experiments, radiation detection, and as astronomical photodetectors. A CCD detects incident photons impinging on its photosensitive region by measuring the number of electron-hole pairs created via the photoelectric effect in a semiconductor substrate, typically silicon. Charge is then transferred along pixels in a column by varying clock voltages until they are shifted to the last row, the serial register, where charges are shifted horizontally to a sense node MOSFET for charge measurement. Since their first application in astronomical imaging in 1976, CCDs have been the most widely used sensor in astronomy for both ground and space based telescopes because of their high quantum efficiencies ($>90\%$) in the optical to near-IR wavelengths, excellent linear response, and outstanding low-light performance \citep{2001sccd.book.....J}.

When performing precision astronomical measurements, multiple sources of noise including shot noise from the source or sky background, dark current, and on-chip amplifier read noise hamper the effective sensitivity of the detector. While imaging observations on the ground can be designed to be shot noise-limited and dark current can be mitigated to negligible levels ($\sim$ 1 e-/pix-hr) by cooling the detector, reducing the read noise is an important aspect that drives improved observational reach in high-resolution spectroscopy \citep{1991PASP..103..122N}. In particular, optical spectroscopic surveys have exposure time restrictions and for shorter wavelengths where the sky is fainter, the readout noise can dominate the overall noise budget \citep{2020SPIE11454E..1AD}.  Current state-of-the-art scientific CCDs used by astronomical facilities are constrained to approximately 2.5 e- of read noise. As observations in the low-background regime push toward fainter, more distant objects, there is a need to overcome this barrier to achieve nearly quantum-limited detecting capability for next generation spectroscopic measurements from the ground \citep{2022arXiv220903585S} as well as space-based spectroscopy and imaging \citepalias{2021pdaa.book.....N}.

One of the central challenges for high-fidelity spectroscopic surveys is balancing the mapping speed with target depth. The mapping speed is driven by many factors including exposure times to reach required signal-to-noise ratios (S/N), and overheads such as slewing and readout time which must be optimized as part of the survey strategy \citep{2023AJ....166..259S}. The Dark Energy Spectroscopic Instrument (DESI) is collecting 40 million galaxy and quasar redshifts to constrain the dark energy equation of state and the growth history of the Universe by probing the baryon acoustic oscillation (BAO) scale and clustering anisotropies \citep{2016arXiv161100036D}. A typical dark time DESI exposure time ranges from 1000 to 1800 s to successfully obtain redshifts in faint target samples \citep{2023AJ....165..144G, 2023ApJ...944..107C, 2023AJ....165...58Z}. If detector noise is minimized to the single-electron level, the S/N can be increased up to 40-70\% at the nominal DESI exposure times at 400 nm, a significant improvement for resolving the redshifted Lyman-$\alpha$ (Ly$\alpha$) forest of distant quasar spectra, which is critical for understanding the intergalactic medium fluctuations carrying BAO signatures. For the same S/N, defined to be a minimum of 1 per angstrom in the Ly$\alpha$ forest for DESI, this would equate to about a 50\% reduction in the required exposure time. Decreasing exposure time opens the flexibility for higher cadence observations and a viable path forward for increasing mapping speeds of future surveys, and thus greater volumes, higher densities, and wider redshift ranges. Improved detector technology is one crucial component for increasing the mapping speed of spectroscopic surveys by an order of magnitude to achieve a Stage-V spectroscopic survey (Spec-S5), which would extend the accessible redshift range to $2 < z < 5$ with a projected $\sim 10^8$ measured redshifts in a new, unexplored discovery space \citep{2022arXiv220904322S,2023P5}.

As a bridge toward the next generation spectroscopic instrument, a potential upgrade of DESI in consideration will double the redshift yield currently planned for the main DESI survey by probing intermediate high-redshift targets around $z > 2$, deep within the matter-dominated era. Such an instrument would pilot CCDs capable of sub-electron read noise in preparation for Spec-S5. This would also maximize scientific returns from Rubin Observatory Legacy Survey of Space and Time (LSST) imaging surveys through cross-correlation with wide-field (thousands of square degrees) spectroscopic samples selected from order $10^8$ planned Spec-S5 targets for photometric redshift calibration \citep{2019BAAS...51c.363M, 2022arXiv220903585S}.

Many other science cases also stand to benefit from ultra-low noise CCDs, notably in space-based facilities where sky background is absent. Identifying biosignatures in light reflected or absorbed by the atmospheres of extrasolar planets is the scientific driver for the next flagship NASA space observatory, the Habitable Worlds Observatory (HWO), and requires developing extremely high-contrast coronagraphy or occulter techniques \citepalias{2021pdaa.book.....N}. Key spectral features potentially indicative of an ``Earth-like" planet include water vapor and molecular oxygen lines at 940 nm and 760 nm, respectively, and sensitivity in this wavelength range would be considered a minimum observational requirement \citep[e.g.,][]{2014PNAS..11112634S}. However, even with a perfect coronagraph, photon count rates are expected to be around only $<0.5$ photons/s/m$^2$ for broadband imaging \citep{2010exop.book..111T, 2016JATIS...2d1212R}. For integral field spectrographs where an already low flux is further dispersed over a large pixel area by additional optics, the count rate drops to a few photons per hour per pixel, translating to a read noise requirement $\lesssim 0.1$ e- rms/pixel \citep{2015ApJ...808..149S, 2016JATIS...2d1212R}. A fast readout from a 2k $\times$ 2k four-corner amplifier sensor of $< 1$ min in addition to exposure time limits are also necessary to prevent high-energy protons in cosmic rays from corrupting a significant ($\gtrsim 10\%$) fraction of pixels \citep{2019SPIE11115E..1ER}. This would also greatly benefit high-cadence transient searches which aim to find and quickly provide follow-up spectroscopic observations throughout the light curve evolution of these objects. Consequently, photon-counting capabilities with high readout rate, excellent response across the optical to near-infrared wavelengths, and robust radiation tolerance are essential for candidate detectors identified for such an application.

The multiple-amplifer sensing (MAS) CCD offers a desirable array of characteristics that appeal to the criterion for future ground spectroscopy and space imaging \citep{2023AN....34430072H}. Using a modified architecture of the Skipper CCD, described in detail in \S\ref{sec:skipper2mas}, the MAS CCD couples fast readout with the sub-electron noise floors enabled by the Skipper CCD. As the latest iteration of legacy thick, $p$-channel devices, the MAS CCD builds on the precedence of fully depleted sensors developed at Lawrence Berkeley National Laboratory \citep{2003ITED...50..225H,2023AN....34430072H} for astronomical applications from SNAP and the Dark Energy Camera (DECam) to the Keck Telescope and DESI. These devices feature high quantum efficiency into the near-IR and radiation resilience due to their $p$-channel design \citep{ROE2007526, Dawson:2007yi, 10.1117/12.856818, 2017JInst..12C4018B}. In this work, we instrument, test, and optimize a prototype MAS CCD using DESI electronics to realistically determine its feasibility as an astronomical detector for next generation spectroscopy, especially focusing on a DESI upgrade as a pathfinding instrument. Complementary tests of a 16 channel MAS CCD are presented in \citep{2024arXiv240519505L}, but in our study, we focus on the performance of the MAS CCD in the context of astronomical spectroscopy, particularly for DESI and Spec-S5.

This paper is outlined as follows: in section \S\ref{sec:skipper2mas}, we place the MAS CCD in the context of Skipper CCD developments and describe its basic properties; in section \S\ref{sec:readout}, we give an overview of our experimental setup for CCD testing and the controller electronics; in section \S\ref{sec:noise}, we demonstrate the noise performance of the MAS CCD under various readout settings; in section \S\ref{sec:characterization}, we show key characterization measurements for the device; in section \S\ref{sec:plans}, we present ongoing and future plans with the MAS CCD in light of our results; and finally, we conclude with our main findings in \S\ref{sec:summary}.

\section{From Skipper to MAS CCDs} \label{sec:skipper2mas}

\begin{figure*}
    \centering
    \includegraphics[scale=0.53]{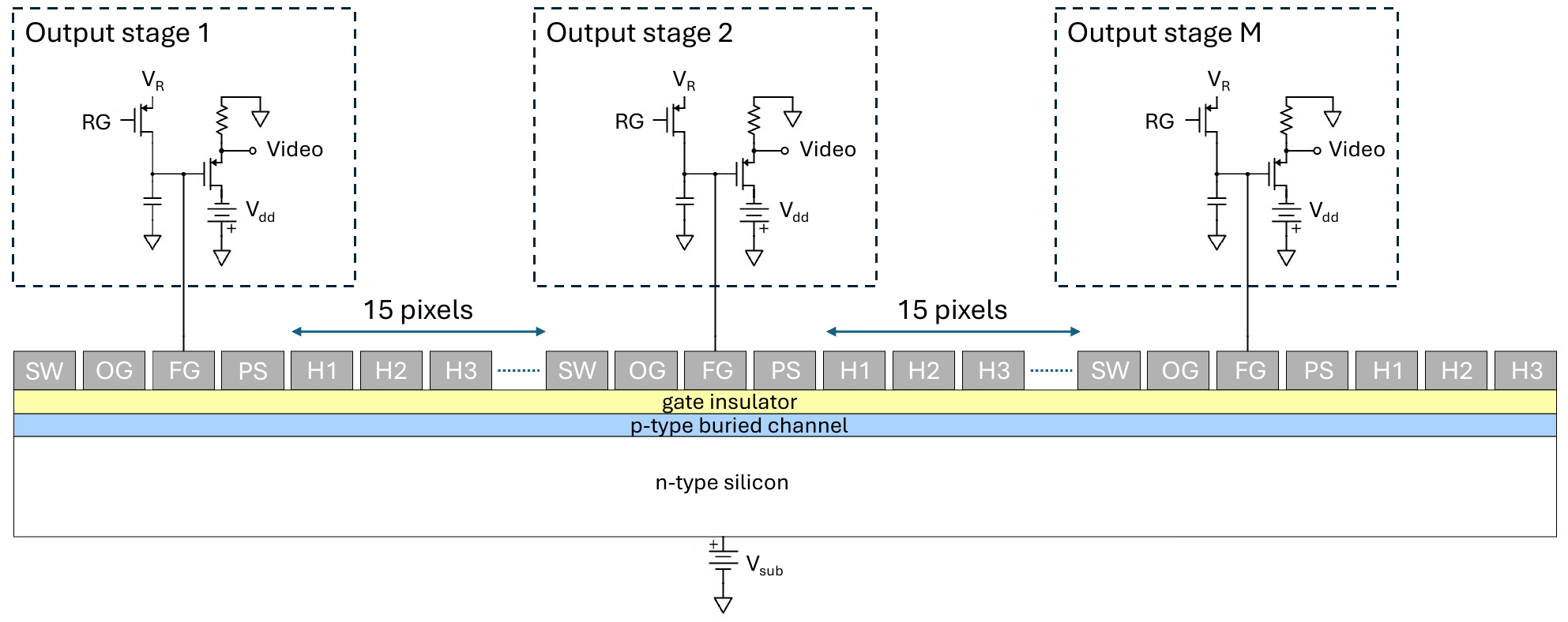}
    \caption{Cross-sectional schematic of the MAS CCD readout structures. For the 16 channel MAS CCD, these structures are repeated 16 times. Each output stage is identical, consisting of two transistors in the floating gate (FG) area: reset with the reset gate (RG) and output. Each output stage is separated by 15 inter-amplifier pixels ($k_{\rm inter}$) with horizontal clocks: H1, H2, and H3 for three-phase clocking. \new{The number of inter-amplifier pixels is set by the spacing of the bond pads, which have been conservatively designed in this prototype to more easily accommodate wire bonding.} The other gates in the readout structure are the summing well (SW), output gate (OG), and the pixel separation (PS). In this work, we study the device with the output gate biased instead of clocked.}
    \label{fig:masschematic}
\end{figure*}

The Skipper CCD circumvents the $1/f$ readout noise limitation by employing a floating-gate amplifier where a charge packet can be independently and non-destructively sampled multiple times \citep{Janesick_1993}. First proposed by Janesick for low-signal-level imaging and spectroscopy, the repeated measurements of each charge packet allows the readout noise to be reduced to the sub-electron level in the first tests \citep{1990ASPC....8...18J}. Modern implementations of the Skipper CCD designed at Lawrence Berkeley National Laboratory (LBNL) have been demonstrated to reach a readout noise as low as 0.039 e- rms/pixel \citep{2017PhRvL.119m1802T,2021JATIS...7a5001C}. This extreme sensitivity has allowed Skipper CCDs to be successfully exploited for an array of broad applications in particle physics, including direct dark matter detection, low-energy neutrino experiments, and fundamental physics studies of silicon \citep[e.g.,][]{2018PhRvL.121f1803C,2019PhRvL.122p1801A, 2021JPhCS2156a2115N, 2022arXiv220210518A, 2022PhRvD.106i2001N, 2021JInst..16P6019A}. Their demonstrated success in this arena and heritage as an outgrowth of conventional silicon CCDs has attracted interest in their adoption for cosmological surveys and direct imaging of exo-Earths from space \citep[e.g.,][]{2017SPIE10398E..0HC, 2019JATIS...5b4009S, 2020SPIE11454E..1AD, 2022SPIE12180E..65R, 10.1117/12.2629475,2024arXiv240519418S}.

While the Skipper CCD provides unprecedented sub-electron resolution, it suffers from a readout time penalty directly proportional to the number of reads per pixel. Unlike a conventional CCD where charge is passed to a floating diffusion amplifier once and then drained, the averaging of many statistically independent measurements in Skipper CCDs allows the noise to be reduced as the inverse square root of the number of individual reads. As the number of samples increase, the resulting readout noise is lowered but at the cost of readout time because only a single pixel at a time can be transferred between the sense node and floating gate in the output stage. For a survey such as DESI, this is projected to require over a prohibitive 15 minutes of readout time to reach a resolution of 1 e- rms/pixel for a DESI-like format (4k $\times$ 4k pixel) four-corner readout CCD \citep{2017PhRvL.119m1802T,2024PASP..136d5001V}. The problem of readout time is further exacerbated for time-domain surveys, particularly with fast-evolving transients where early-time observations conducted with high-cadence searches are crucial for characterizing the progenitors of these objects.

First developed as part of a U.S. Department of Energy Quantum Information Science initiative, the MAS CCD was designed to address the trade-off between reaching sub-electron levels of read noise and extended readout time that is a key drawback for single-amplifier Skipper CCDs \citep{2023AN....34430072H}. This is accomplished by implementing a chain of floating-gate output amplifiers, interspersed along an extended serial register. A schematic of this readout structure is illustrated in Fig. \ref{fig:masschematic}. Charge packets pass through and are sampled non-destructively by each output stage. Each charge packet is thus repeatedly measured to reduce noise in a similar approach to the Skipper CCD, but allows the charge to be continuously transferred to the next amplifier for the subsequent charge packet to be measured. This architecture permits sampling $M$ charge packets simultaneously for $M$ floating-gate amplifiers to effectively parallelize readout, thereby decreasing the readout time. Thus, the effective noise per pixel after averaging $N$ uncorrelated charge measurements for $M$ amplifiers is
\begin{equation} \label{noiseequation}
\sigma = \frac{\sigma_1}{\sqrt{N \cdot M}}
\end{equation}
where $\sigma_1$ is the noise per pixel for one charge measurement for an amplifier. The theoretical read time for a MAS device with $m$ number of rows and $n$ columns as described by \citet{2023AN....34430072H} is
\begin{equation}
t_{\rm read} = m (t_n + t_V)
\end{equation}
where $t_V$ is the vertical clocking row shifting time and $t_n$ is the time required to read a $n$ pixels in a row, equivalent to the total number of columns of the active area. The time to read a row  $t_n$ is given by
\begin{multline}
    t_n = k_{\rm ext} t_{\rm shift} + M(k_{\rm inter} t_{\rm shift} + Nt_{\rm read}) \\ + (n-1)(t_{\rm shift} + Nt_{\rm read})
\end{multline}
where $k_{\rm ext}$ are the number of extended serial pixels, $k_{\rm inter}$ are the number of pixels between amplifiers, $t_{\rm shift}$ is the time to shift charge one serial pixel, and $t_{\rm read}$ is the time to perform one charge measurement in the sense node \citep{2023AN....34430072H}. The first two terms are equal to the time to read the first pixel completely through all MAS amplifiers in the serial register and the third term is the time it takes for all subsequent pixels to pass from the penultimate to the final amplifier, since after the first pixel has read out, the second pixel has been read out by all but the last amplifier. Incorporating multiple output stages also introduces the benefit of amplifier redundancy against amplifier failures that would otherwise put an entire conventional CCD at risk, since charge can simply flow through non-functioning output stages.

Although the MAS output stages are also capable of measuring charge by ``skipping," or transferring charge between the sense node and summing well repeatedly to increase the number of charge measurements per amplifier $N$ by virtue of the floating-gate amplifier structure \citep[e.g.,][]{10521851}, we focus on the fastest readout mode for the sensor where $N=1$, measuring each charge packet exactly once per MAS amplifier. A single-sample per amplifier readout is desirable in light of readout time requirements for large-format, scaled-up variants of the MAS CCD that satisfies mapping speed and noise requirements ($\sim$ 1 e- rms/pixel) of surveys. A fast readout also minimizes pixel dwell time inherent in ``skipping" when charge must sit in the active area as $M$ pixels are being measured by the $M$ output stages, downtime that not only cannot be used for an additional exposure, but also increases the fraction of pixels contaminated by cosmic rays.

\subsection{Device properties}
The test device was fabricated by Teledyne DALSA Semiconductor from high resistivity (10 k$\Omega$-cm) $n$-type silicon substrate with buried $p$-type channels and no backside processing or anti-reflective (AR) coating. There are 16 MAS readout channels along the 256-pixel extended serial register. On the opposite end of the serial register is a DESI-like output amplifier which, if connected, can enable the sensor to act as a conventional CCD when charge is clocked in the direction away from the MAS amplifiers. Standard three-phase clocking is employed to read out charge vertically and horizontally, and the serial clocks are similar to that of Skipper CCDs with the exception of one additional pixel separation (PS) gate. The PS gate acts as a potential barrier between floating-gate sense node and subsequent inter-amplifier pixels. This output structure is illustrated in the charge transfer diagram in Fig. \ref{fig:masschematic}. In the final output stage, a dump gate (DG) takes the place of the PS gate, and is structurally similar to the Skipper CCD output stage. Since the serial register is only one edge of this prototype detector, there is only one set of vertical clock lines, and the pads on the CCDs on both sides of the device that correspond to these vertical clocks are connected to the same set of clock drivers. The sets of serial clock voltages in the MAS register are also tied together, so all MAS output stages have the same set of clock voltages. The device has a format of $1024 \times 512$ pixels with 15 $\mu$m pixels and since the sensor was not thinned, the thickness was 650 $\mu$m and was illuminated on the frontside. These properties are summarized in Table \ref{table1}.

\begin{deluxetable}{ccc}
\label{table1}
\tablecaption{Summary of characteristics and operating conditions of the MAS CCD tested in this work.}
\tablehead{\colhead{} & \colhead{Value} & \colhead{Units}}
\startdata
Active area       & 1024 $\times$ 512      & pixels \\
Pixel size        & 15 $\times$ 15         & $\mu$m \\
Pixel pitch       & 15              & $\mu$m \\
Device thickness  & 650             & $\mu$m \\
Output stages     & 16 MAS + 1 DESI & \\
Temperature       & 143             & K \\
Substrate voltage & 40              & V \\   
\enddata
\end{deluxetable}

\subsection{Detector packaging}
\begin{figure}
\centering
\includegraphics[scale=0.16]{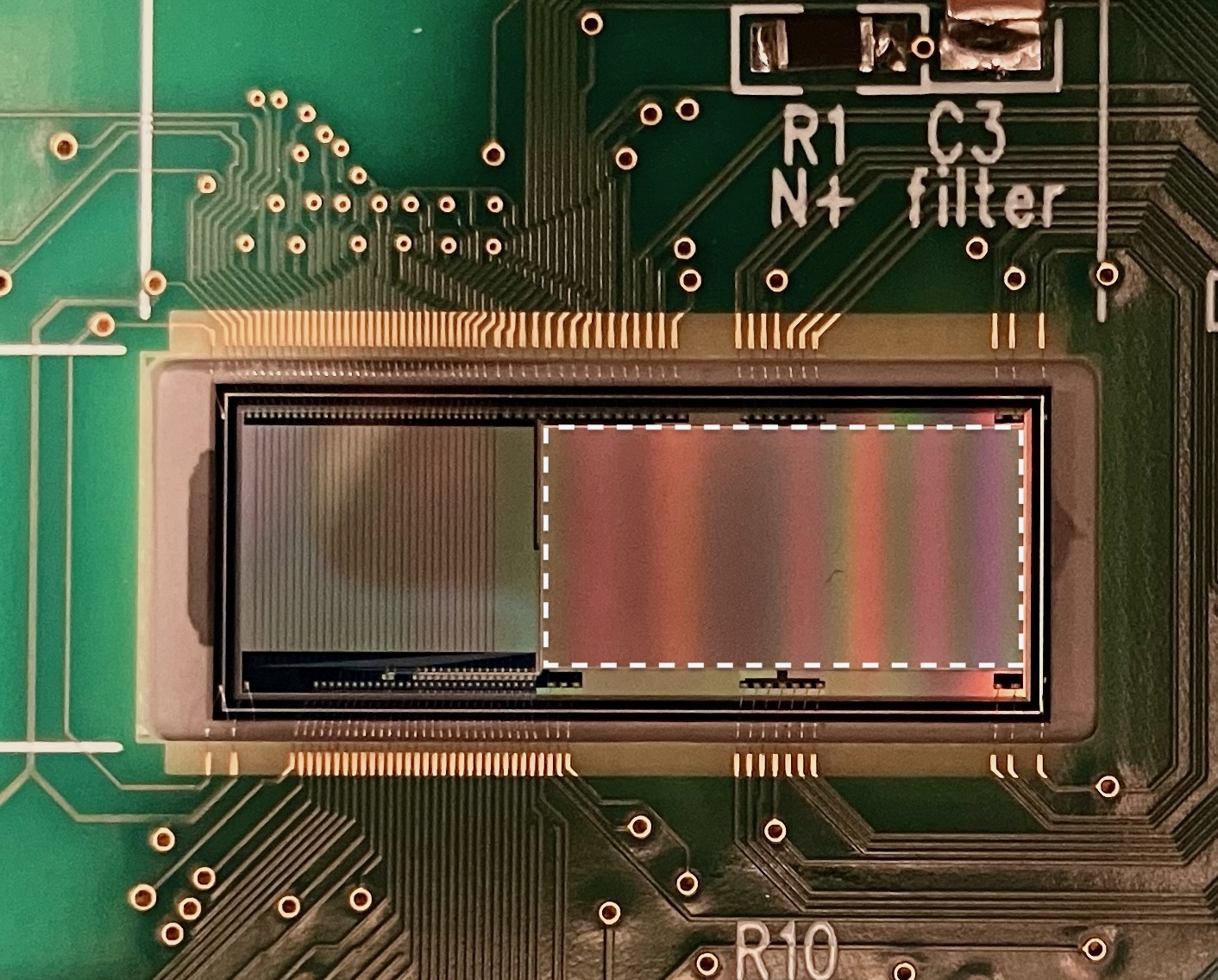}
\caption{A prototype 16-channel MAS CCD packaged and wirebonded onto the carrier board and installed in the cryostat. The active photosensitive pixels are outlined in the dashed line and the serial register is along the left edge. The area to the left of the active region is unused silicon. The MAS in-line amplifiers extend from the bottom left corner of the active area, where the 16 bonding pads corresponding to the 16 amplifiers are visible.}
\label{fig:maspicture}
\end{figure}

A ``picture frame" format carrier was designed for the CCD with an opening slightly larger than the size of the sensor centered on a printed circuit board (PCB). The carrier board includes a 1 $\mu$F and 4 M$\Omega$ filter in parallel visible in Fig. \ref{fig:maspicture}, which helps stabilize the output voltage of the video transistor \new{by stabilizing the potential of the n+ guard structures }\citep{5280320}. A square silicon substrate was bonded to the back of the PCB, covering this opening. The backside of the CCD was then carefully bonded onto the substrate with epoxy through the opening on the front side of the PCB, then baked to cure. The CCD pads were then wirebonded with the PCB. A ribbon cable was used to connect this package to the dewar vacuum feedthrough. The wirebonded detector is pictured in Fig. \ref{fig:maspicture}.

\section{Experimental apparatus and readout electronics} \label{sec:readout}
The CCD was mounted in a dewar cooled by liquid nitrogen under vacuum at about $10^{-4}$ torr using a diaphragm roughing pump and turbomolecular pump to maintain high vacuum. Using a Lakeshore 336 PID temperature controller, the CCD was operated at 143 K as measured by a temperature sensor positioned at the cold finger on the rear of the CCD package. Since liquid nitrogen dispenses at 77 K, the CCD operating temperature was reached by mounting the CCD package on a cold plate that is partially thermally isolated from the cold finger with copper foil bridging the thermal contact between the two components. Power resistor heaters were placed on the cold plate to regulate the the temperature. The dewar enclosure has a retractable source holder arm on the inside next to a fused silica window on the latched dewar chamber opening on the front where the CCD is installed. This window provides an optical path for illumination. Optical elements, including an iris diaphragm, focus tuning, and LED source with a built-in lens holder in front of the diode, are attached onto a baffle connected to the dewar window. Different lenses can be placed in front of the LED to project mask patterns onto the CCD. The LED source is connected to a custom external controller that can change the light intensity, synchronize the LED with a specified exposure time to act as a shutter while operating the CCD, or provide manual control. The dewar including the LED projection system was covered with an opaque cloth to minimize stray light entering the dewar especially on the front side of the dewar opening and connection points between components outside the dewar.

The dewar vacuum feedthrough links the ribbon cable from the CCD package to a dewar side board connected to the readout electronics. The CCD control is performed with an expanded version of the DESI front-end electronics (FEE) system to enable the simultaneous digital correlated double sampling (CDS) readout of the 16 output channels on the sensor. The new electronics developed, called the Hydra, runs as a set of four synchronized DESI FEE systems in a leader-follower configuration, consisting of a video processing board and power conversion board for each of the four FEE sub-units and a single clock driver board for the master FEE. The video board serves to process and digitize four channels of CCD video signal using a 100 MHz 16-bit analog-to-digital converter (ADC). The clock board provides the clock signals with a range $\pm$ 12 V for the CCD with a 40 channel 16-bit digital-to-analog converter (DAC). Each component of the FEEs within the Hydra sits on a motherboard with an FPGA which is the timing engine and reads ADC data output from the video board, and CDS data is sent to a MicroBlaze processor accessible by a TCP/IP communication interface. Data products are written out in standard FITS image format. Four main operating modes are defined for the CCD in our setup: exposing, readout, idle, and cleaning. A clear is performed prior to exposures and is intended to clear remnant charge from the active area by completing a readout sequence without reset or signal sampling. Idle mode continuously clocks the CCD between exposures and is stopped during a set exposure time and readout.

An aluminum enclosure with four cooling fan exhaust ports was assembled to shield all the electronics from the environment. To reduce interference within the casing, thin aluminum plates were installed between each FEE subunit. Each FEE could be power cycled by individual switches which were installed on the exterior of the entire enclosure.

\section{Noise performance} \label{sec:noise}
For a CCD image with zero exposure time, pixel values are the sum of the dark current accumulated during the readout time, the bias level DC offset, occasional cosmic ray signal, and amplifier read noise. As a functional definition, we refer to dark current in this work as the signal detected when the CCD is not exposed to light, and note that the thermal leakage current is only one component among other dominating components including stray light and thermal photons from the dewar walls; we do not attempt to separate these components in this study. Extended strips of virtual pixels along the image borders created by overclocking the CCD array comprise the overscan region, where the distribution of pixel values reflect the read noise and bias level. \new{A strip of 27 pixels called the prescan is also created along one edge of the CCD. These pixels are on the die with no associated column of active area pixels located between the first active column and the summing well of the first MAS output amplifier.} We note that these pixels also contain other components such as the dark current and cosmic ray contributions accumulated during the readout of the rows and columns. Overscan subtraction for the entire image including the virtual pixels corrects for the bias shift of all pixels by taking the sigma-clipped mean\footnote{The sigma-clipped mean is an iterative process consisting in rejecting outliers (at $N$ standard deviation from the mean), and re-evaluating the mean and standard deviation of the remaining pixel values (e.g., \href{https://docs.astropy.org/en/stable/api/astropy.stats.sigma_clip.html}{astropy implementation}).} of the overscan region and subtracting this from each pixel. During testing, we found this yields similar results to a line-by-line overscan correction where the sigma-clipped mean pixel value for each row of the serial overscan is subtracted from each physical and virtual pixel in the same row. We apply iterative sigma clipping which rejects outlier pixel charge contaminated by cosmic rays from the analysis. The standard deviation derived from fitting the resulting overscan pixel distribution after overscan subtraction and sigma clipping corresponds to the read noise. 

\begin{figure}
\centering
\includegraphics[scale=0.372]{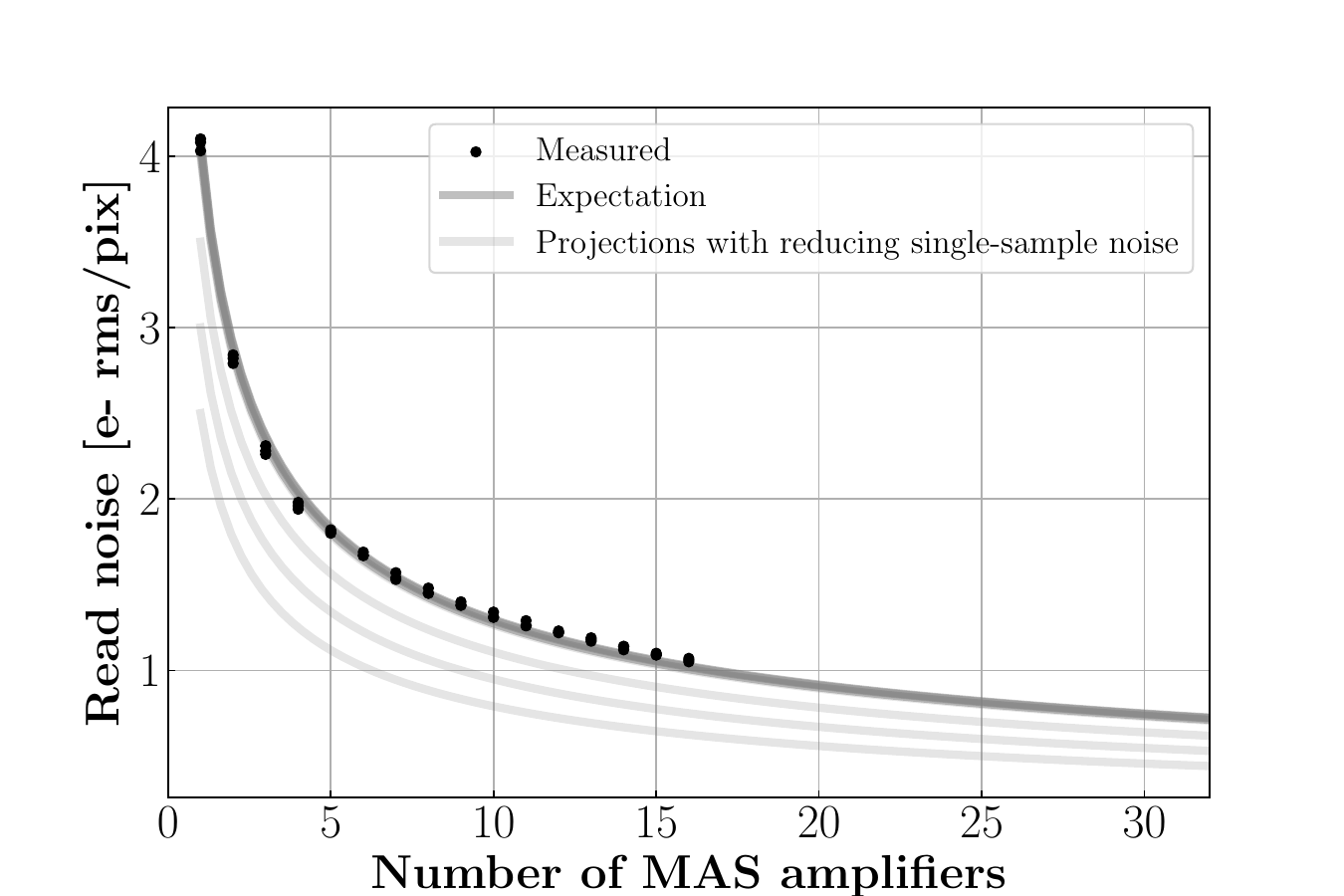}
\caption{Read noise for a single sample read per amplifier is measured for the data points in comparison with the expected reduction curve given the starting read noise at the first amplifier. Shown in the gray curves are possible future optimizations in amplifier noise with a 32 channel design which could enable deeply sub-electron noise with only a single-sample read per amplifier. These extrapolations are based on the possible single-sample Skipper CCD noise measurements demonstrated at 3.5 e- rms/px and if each amplifier had an individual noise at this level \citep{2017PhRvL.119m1802T}. The most optimistic scenario where amplifier noise is 2.5 e- rms/px is based on currently DESI red/NIR CCDs operating at this level and with long pixel integration times on Skipper CCDs \citep{2021JATIS...7a5001C}. Different readout mechanisms enabled by multiple amplifiers in the MAS CCD and new low-noise electronics may enable single-amplifier read noise to reach the 2.5 e- noise floor achieved by Skipper and DESI CCDs.}
\label{fig:noisereduction}
\end{figure}

\begin{figure*}
\centering
\includegraphics[scale=0.43]{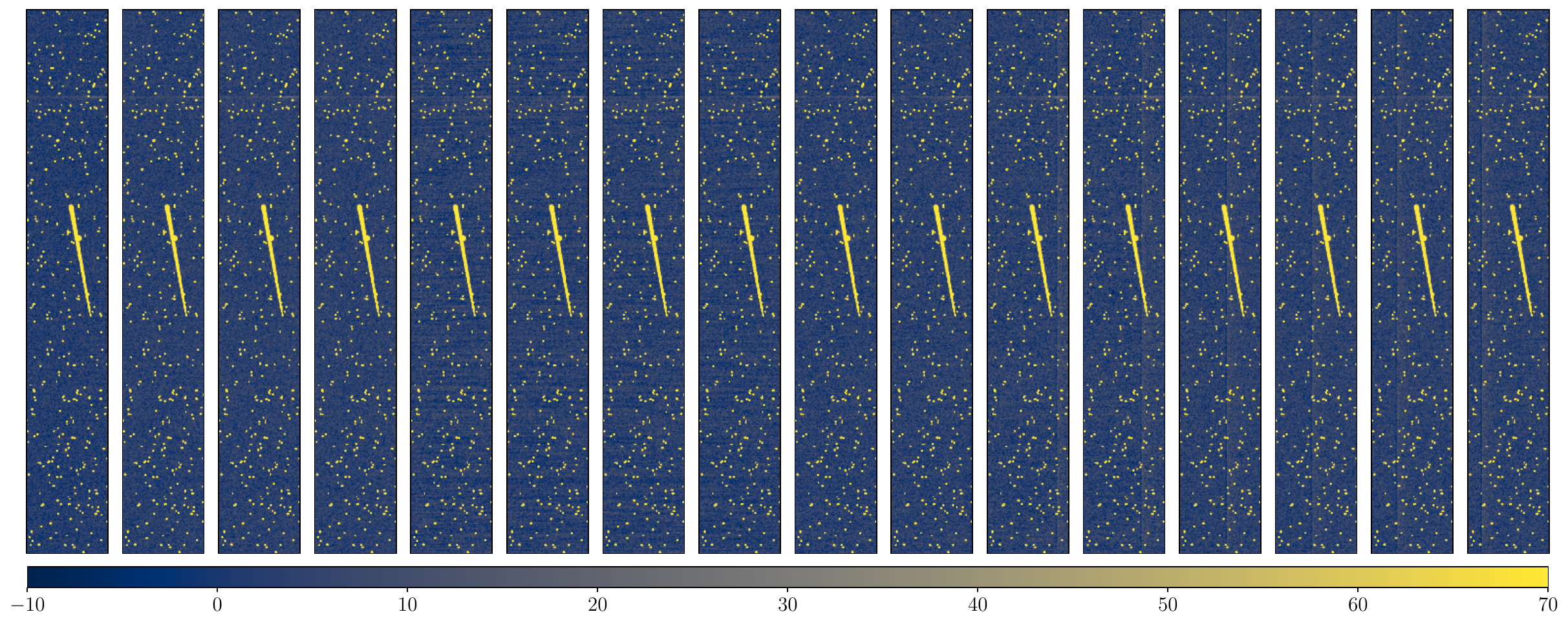}
\caption{Frames from each of the 16 output amplifiers of the MAS CCD under X-ray exposure (point-like clusters) after alignment correction. The straight track is a signature of a cosmic ray muon.}
\label{fig:masframes}
\end{figure*}

Because the MAS CCD produces the same number of images as the number of output amplifiers for a single sample per amplifier in an exposure, in our case 16, these image frames can be used to measure the noise of its corresponding amplifier. The 16 images are then averaged to measure the combined noise for the multiple charge measurements made per pixel. The effect of averaging on noise as a function of the number of MAS amplifiers is shown in Fig. \ref{fig:noisereduction} in comparison to the theoretical expectation from Eq. \ref{noiseequation}. However, before the signal is averaged, the frames must be regridded to account for the 16 inter-amplifier pixel shifts, where each image starting from the second amplifier in the MAS chain is shifted 16 pixels serially forwards compared to the image output from the preceding amplifier. This correction was implemented by shifting images from the second amplifier onwards to share the pixel grid of the image from the first amplifier. These steps of shifting, signal averaging, and overscan subtraction were applied to each image as part of our standard preprocessing. \new{An example of the 16-amplifier MAS CCD image output is shown in Fig. \ref{fig:masframes}.}

\subsection{Gain calibration} \label{gaincalibration}
\begin{figure}
\centering
\includegraphics[scale=0.39]{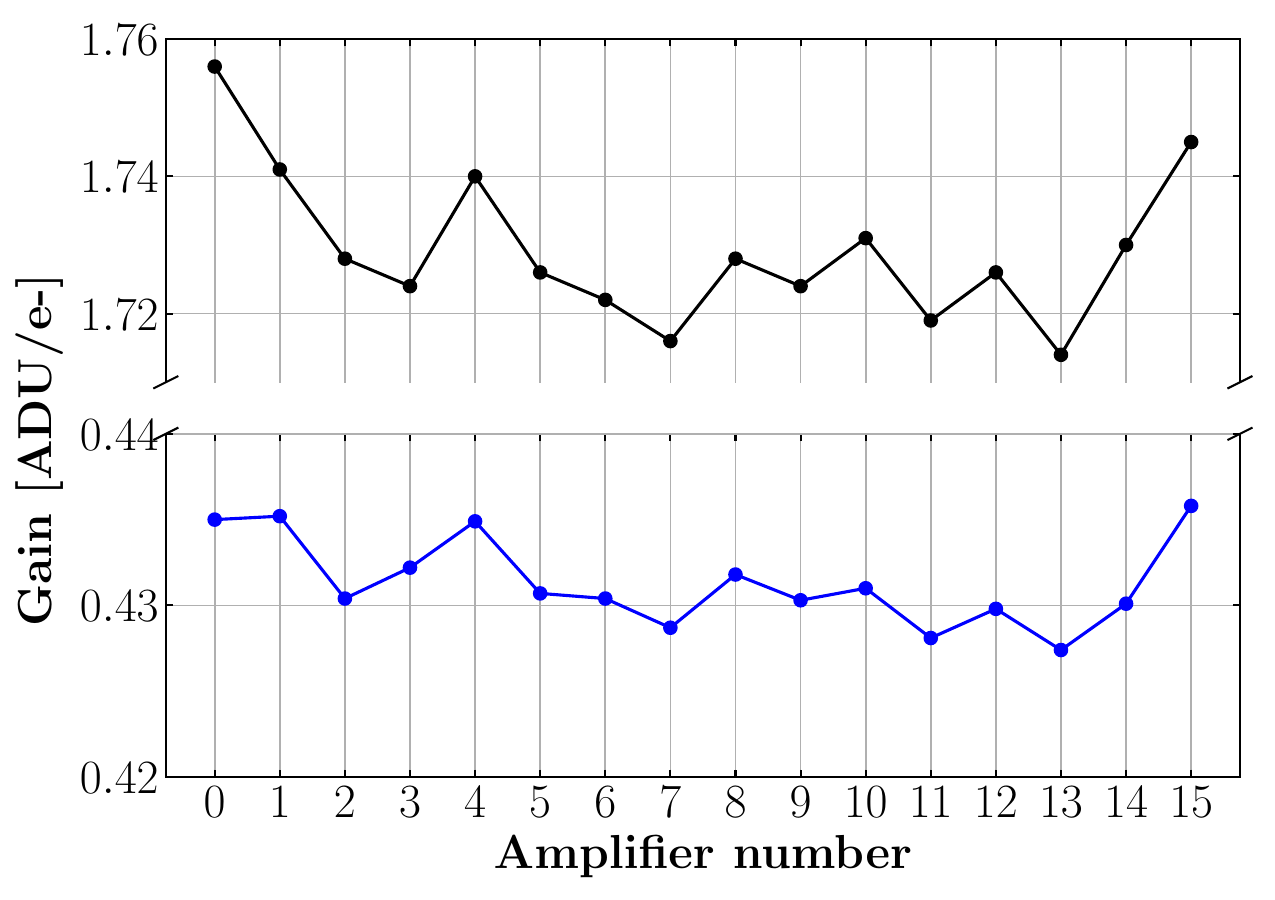}
\caption{Conversion gains for each of the active MAS amplifiers at ADC gain 5 (black) and ADC gain 3 (blue).}
\label{fig:amplifiergains}
\end{figure}

The detector gain is a conversion factor between counts or analog-digital units (ADU) from the ADC output and number of electrons the CCD recorded, which we measure with the commonly used technique of stimulating the CCD with 5.9 keV X-rays from the radioactive decay of a $^{55}$Fe source. Since the quantity of electron-hole pairs generated in a pixel per X-ray photon absorbed in the silicon bulk is proportional to the X-ray energy and known, the gain can be derived from the measured location of the 5.9 keV Mn K$\alpha$ peak in the spectrum \citep{2001sccd.book.....J}. After obtaining five second exposures \new{where the front side of the CCD was illuminated by the 5.9 keV X-rays}, the X-ray hits were extracted to build a histogram of the charge distribution. \new{This detection of X-ray hits was performed with a standard source extraction algorithm which identifies sources with signal above the image background level \citep{Barbary2016}. This algorithm also performs conventional aperture photometry to extract the flux \citep[e.g.,][]{2006hca..book.....H}. Using a circular annulus with a minimum radius of two pixels and maximum outer radius of five pixels, the sum of the signal in the inner is subtracted by the background signal measured in the annular area between the inner and outer radii. With a histogram of the extracted fluxes,} the Mn K$\alpha$ X-ray peak was then fitted with a simple Gaussian to find the position of the peak in ADU. The conversion gain is obtained as the ratio of this peak position with 1620 e-, the average number of electrons produced by the K$\alpha$ X-ray \citep{1994NIMPA.350..368F}. The gain for each amplifier is shown in Fig. \ref{fig:amplifiergains} for our standard ADC gain setting and a reduced ADC gain used to extend the dynamic range to high signal levels ($\gtrsim 3 \times 10^4$ electrons). Throughout the course of our tests in operating the CCD for one year, we have found the amplifier gains to be highly stable over time.

\subsection{Voltages and noise optimization}
\begin{figure}
\centering
\includegraphics[scale=0.4]{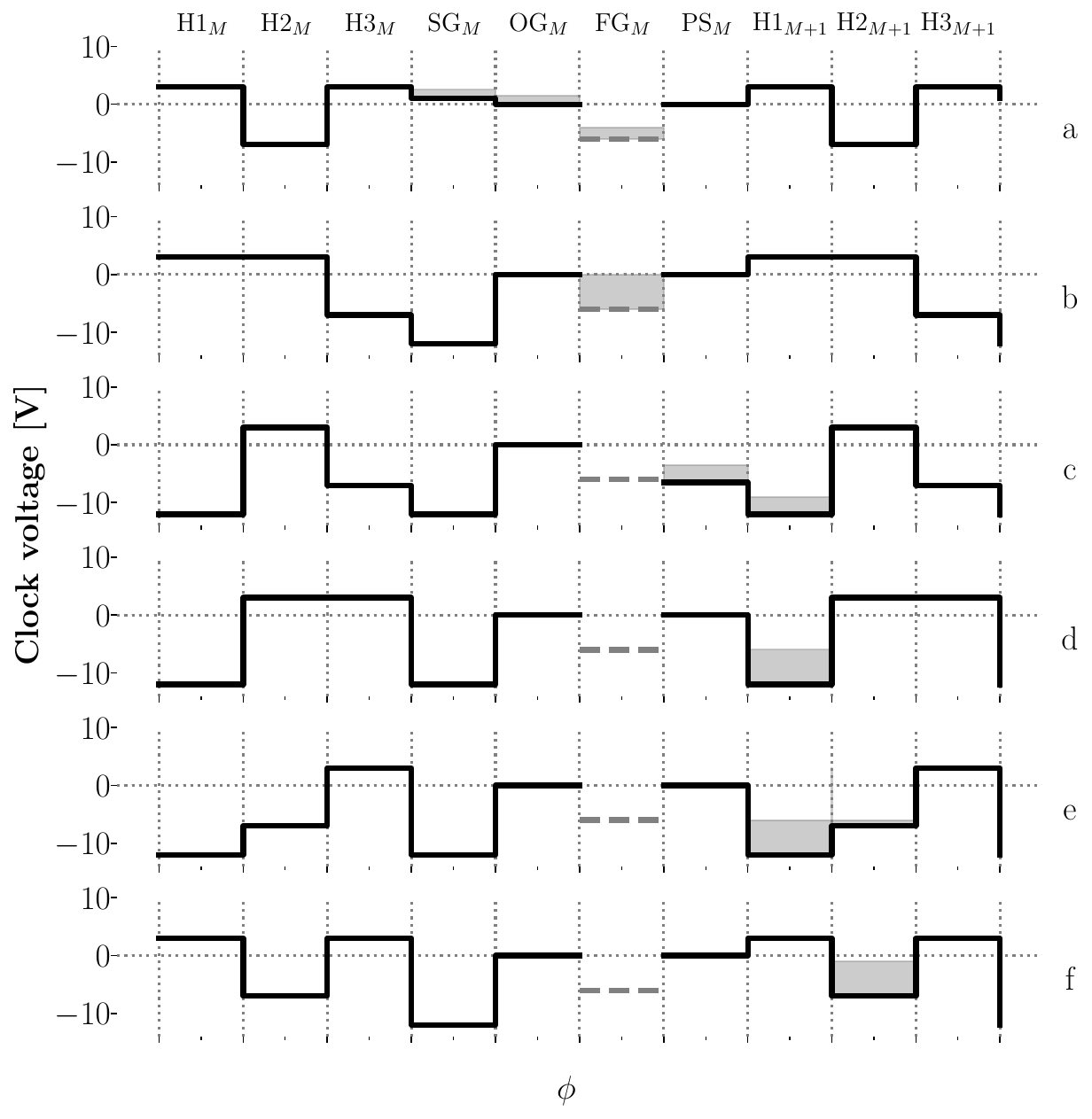}
\caption{A potential diagram of the serial register for a given output stage $M$ is illustrated in our charge transfer diagram, where the floating-gate (FG) sense node denoted by the dashed segment is at an unknown voltage level and is where the CDS is performed to measure the amount of charge present in a pixel. The shaded regions represent the idealized transfer of charge carriers through the phases. The voltage levels corresponding to each phase and state are plotted for our optimized voltages that enable us to reach under the noise of 1.1 e- rms/pixel. In the final output stage, instead of a PS gate, a dump gate (DG) is used which leads to the highly negative $V_{\rm dd}$.}
\label{fig:transferdiagram}
\end{figure}

The CCD was run with a substrate-bias voltage ($V_{\rm SUB}$) of 40 V at which lateral charge diffusion is expected to be smaller than the pixel size. \new{Because the CCD is front side illuminated, both photons from our light source and X-rays are converted to electron-hole pairs close to where the charge is stored and therefore diffusion is minimal \citep{1352164}.} Because we are interested in the performance at the fastest readout configuration where charges are sampled once for each amplification stage (i.e., no skipping), the output gate is biased to 0 V in our setup. The potential diagram with our voltages in the output structure is shown in Fig. \ref{fig:transferdiagram}, where each row of clock sequences represents a state during the readout. The output stage involves the summing gate (SG), output gate (OG), floating-gate (FG) sensing region, and pixel separation gate (PS) and are interleaved by three horizontal clock phases between each output stage. In the last stage, instead of a PS gate, a dump gate (DG) leads to \new{the output drain with a voltage set to -22 V} to remove the charge from the channel. Because the sense node relies on a floating gate for non-destructive charge measurement, the exact potential here cannot be known and is therefore denoted as a dashed segment in Fig. \ref{fig:transferdiagram}. Several clock states are depicted in Fig. \ref{fig:transferdiagram}, where state (a) is where charge enters the sense node, state (b) is where the charge is being measured, and state (c) is where the charge exits the sense node and is shifted away to the next output stage in subsequent states. 

\begin{figure}
\centering
\includegraphics[scale=0.333]{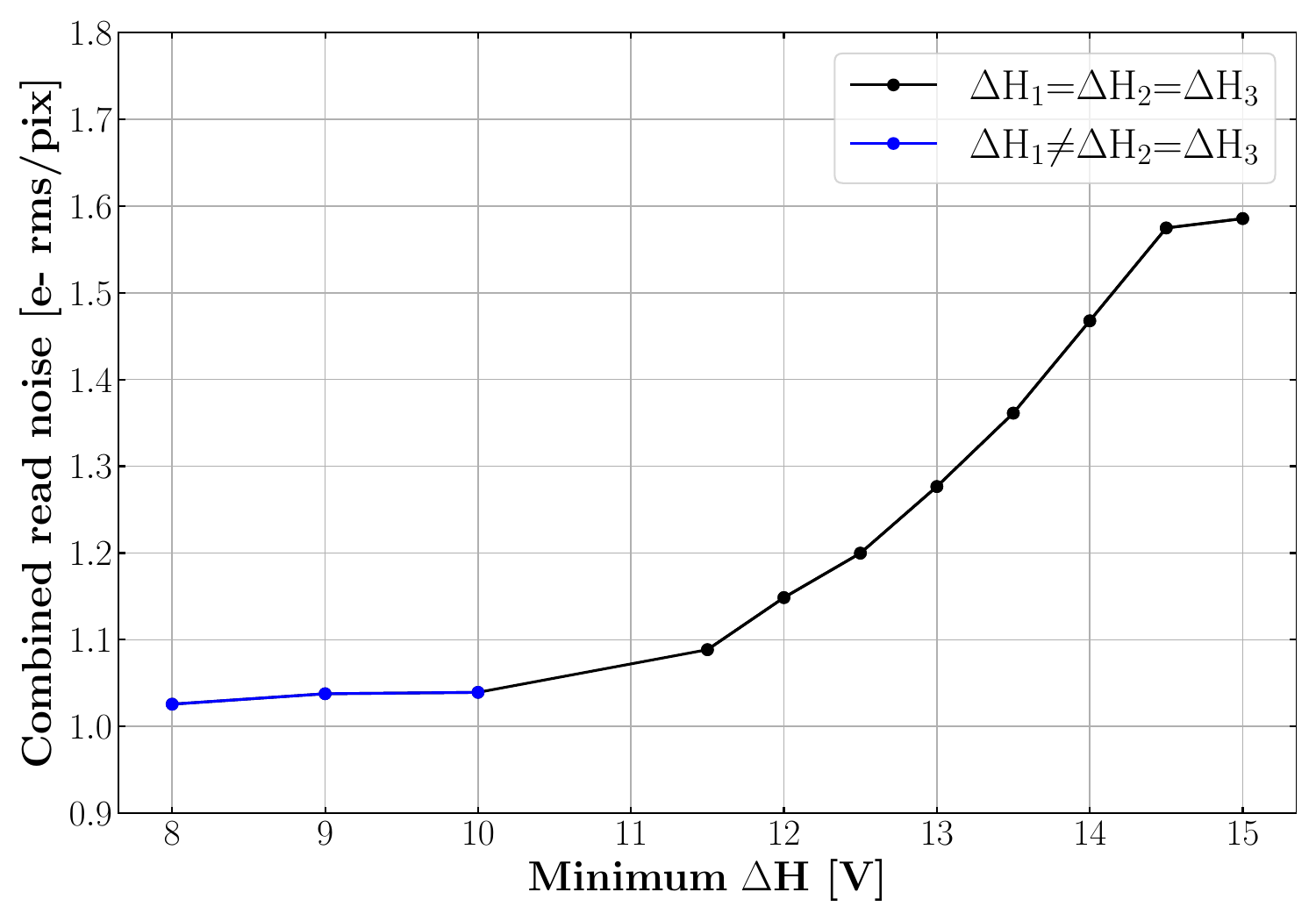}
\caption{Combined read noise as a function of horizontal clock swing amplitudes. For the black data points, the amplitudes are identical for all three horizontal phases while the blue points denote a different amplitude for the H$_1$ clock and are instead plotted against the amplitudes of clocks H$_2$ and H$_3$ which are identical.}
\label{fig:noise_hswing}
\end{figure}

Since the prototype sensor is not divided into quadrants, we used a single set of three vertical clocks for the parallel register and three horizontal clocks for the serial register. Reducing the horizontal clock amplitudes was the most significant driver of noise reduction, allowing the noise floor to be lowered to near single-electron noise. Starting with a 15 V horizontal clock swing, the low voltages of the clocks were gradually increased as shown in Fig. \ref{fig:noise_hswing}. However, there is a limit to the minimum amplitude imposed by charge transfer efficiency between amplifier stages for the first horizontal clock phase, H1. This amplifier-to-amplifier charge transfer inefficiency is linked to the PS gate, where the low voltage of the PS gate must be at least 1 V greater than the H1 low state. When the H1 low state is within 1 V of the PS low state, the reset drain voltage ($V_R$) must be increased to allow potential for the PS low state to increase without causing inter-amplifier charge transfer streaking. Increasing the low voltage of the PS gate was crucial for raising the potential barrier between the sense node and the next stage of horizontal clock phases, when both the PS and H1 clocks are in low states, to prevent charge in one pixel from diffusing \new{back into the floating gate, an effect discussed in \S\ref{actesection}}. For instance, a PS gate voltage at -6.5 V provides a 5.5 V potential well for an H1 at -12 V whereas a PS gate voltage at -8 V only gives 4 V of potential barrier. In this latter case, the well depth formed by this voltage difference is smaller than the parallel clock amplitudes, which have a 5 V amplitude, and becomes a limiting factor for the effective well depth of the detector. At the minimum H1 amplitude with a high of 3 V and low of -8.5 V set globally for all horizontal clocks, the compound read noise reaches just under 1.1 e- rms/pixel with a single-amplifier read noise ranging from 3.68--4.51 e- rms/pixel depending on the amplifier. Although the returns diminish with this approach, it was possible to achieve 1.03 e- rms/pixel of compound noise without incurring charge transfer problems by tuning the second and third horizontal phases independently of the first phase. We interpret this noise to be an effective read noise because its tight correlation with serial clock amplitudes suggests that the noise includes not only intrinsic amplifier noise but also a serial clock-induced charge shot noise component. The serial clock-induced charge was found to a significant contributor to the measured noise at greater horizontal clock heights for a Skipper CCD with a 6000-pixel long serial register but reduces when these clock amplitudes are lowered \citep[cf. Fig. 12 in][]{2024PASP..136d5001V}. Reducing the horizontal clock swings also reduces the full well capacity, which should be balanced against detector noise requirements, discussed in section \ref{sec:fullwell}. 

As part of the DESI FEE standard operating parameters, vertical clocks are wave shaped with no serial clock shaping. We explored adding clock filtering on the horizontal clocks with a 1 nF capacitor and extending the reset delay to accommodate the longer rise times. With the optimized clock voltages without shaping, adding wave shaping caused significant charge transfer inefficiency (CTI) between amplifiers. Increasing the amplitude of the PS gate such that the high voltage is greater than -2 V and the low voltage is at least under -9 V addresses this CTI but since the PS gate influences the allowed neighboring H1 phase, lowering the PS gate low voltage requires H1 to be at least below -10 V. As we have observed, increasing the horizontal clock amplitudes increases the effective read noise and further voltage optimization leads to a read noise slightly above 1.1 e- rms/pixel. Because adding this level of serial shaping increases the trade-off between inter-amplifier CTI and read noise with no noticeable performance improvements, we opted to complete the remainder of our tests without serial clock shaping, noting that it may still be possible to achieve improved performance with a lower capacitance wave shaping implementation.

To estimate the cumulative effects of other noise components such as light leaks in the dewar and parallel register clock-induced charge, zero second exposure bias images were acquired to compare the noise in the overscan region and the active area. The virtual pixels of the overscan region are clocked only in the serial direction for each row. Consequently, differences in the pixel RMS between the two regions are expected to arise from the combination of light leakage and clock-induced charge. The bias images indicate a glow emanating from the top of the parallel register on the ``left hand side" of the device near a corner. The position of this glow directly opposite of where the serial register is located appears to rule out an amplifier origin. To prevent the signal from the glow from being transferred down the parallel register during idle mode \new{where parallel clocks continuously shift charge down the pixel matrix when the CCD is not exposing or reading out}, a stop idle to freeze clocking followed by a clear was implemented before starting the exposure. \new{This clear action clears charge from the active area quickly by shifting out charge in the same clocking sequence as a readout but without reset or signal sampling.} However, because the synchronization of the FEE subsystems requires the followers to be slightly ahead of the leader, there is a practical minimum time delay just under 0.5 s between when the clear is executed and the readout initiated. Thus, we restrict our noise calculations to the bottom 50 rows of the CCD near the serial register and only half the columns on the side towards the overscan where the glow signal has not yet been transferred in the 0.5 s of idle clocking time delay. We find that the difference in the standard deviation of the signal between the active area and overscan is about 0.34 e-. Since this noise quantity includes a combination of the light leaks and total parallel clock-induced charge, we interpret this difference in noise to be an upper limit on the clock-induced charge generated in the parallel register, demonstrating that this effect is not dominating our read noise measurements, making our noise calculation more robust.

\subsection{Integration time}
\begin{figure}
\centering
\includegraphics[scale=0.4]{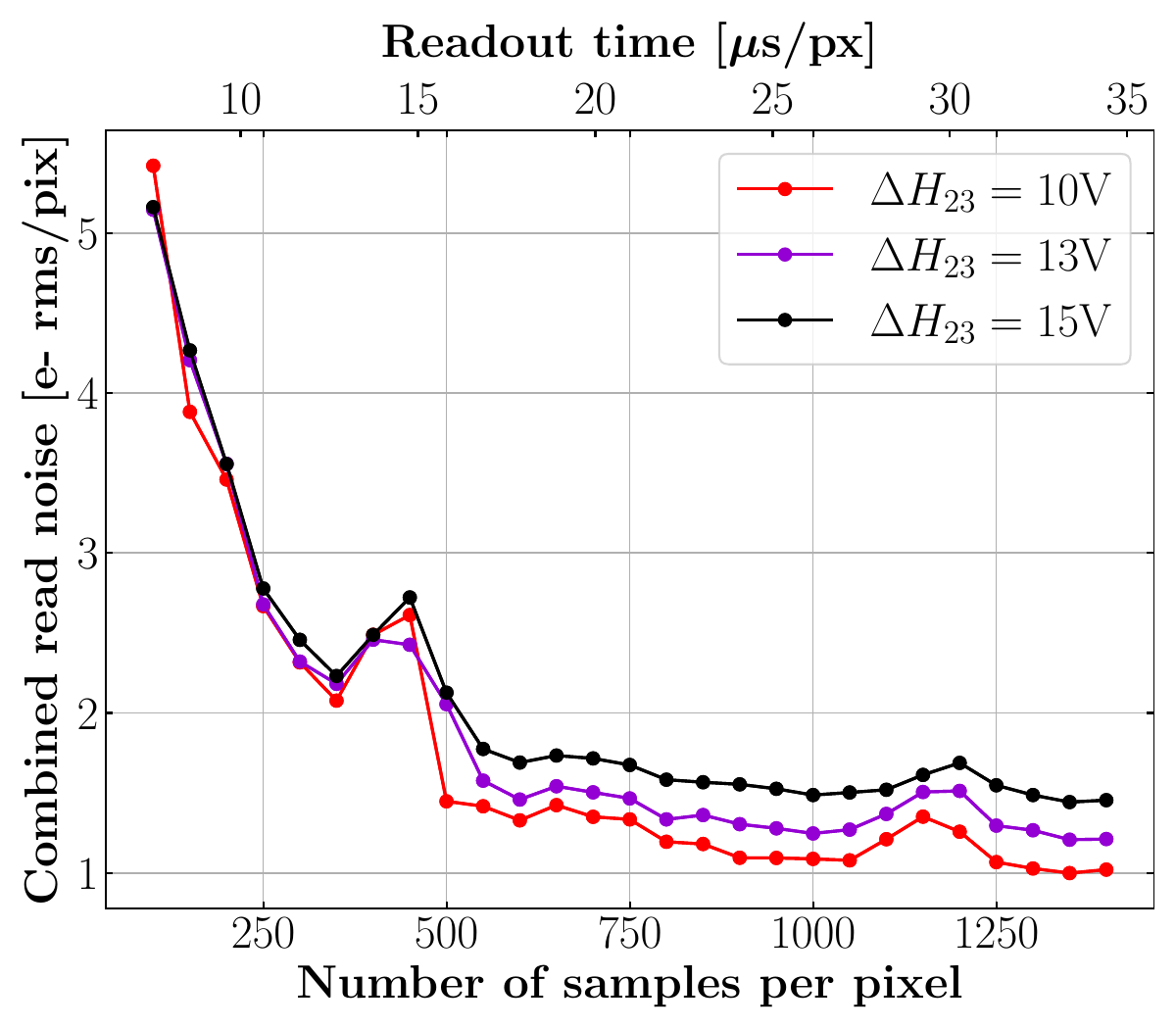}
\caption{Combined noise as a function of the number of samples per pixel (i.e., integration time) taken each at the pedestal and at the signal level. The corresponding readout time is also displayed with the same curve measured at different voltage amplitudes for the H$_2$ and H$_3$ clocks. We operated at 1000 samples per pixel at a readout time of 26.1 $\mu$s/pixel (38.3 kHz) but as shown here, running slightly faster at 900 samples per pixel would give similar noise performance.}
\label{fig:integtime}
\end{figure}

In the CDS readout scheme, the time window in which samples from the CCD video signal are averaged for the signal or pedestal levels is defined as the pixel integration time. This translates to the number of samples per pixel taken at each of these levels, such that with more samples, the noise is expected to decrease but at the expense of the readout time. This is illustrated in Fig. \ref{fig:integtime}, where a larger number of samples per pixel corresponds to a much lower read noise. We read out our device with 1000 samples per pixel, corresponding to a readout time of slightly above 26.1 $\mu$s/pixel, giving a read noise floor of slightly under 1.1 e- rms/pixel. However, the noise does not decrease monotonically and we observe two bumps around 450 and 1150 samples per pixel where the noise increases slightly. This shows that there is a non-white noise component contributing to the overall noise budget, which we attribute to electronics-related pattern noise, particularly since the second bump occurs approximately three half-periods from the first peak, where multiples of half a period constructively increases the pattern noise contribution. We also observe that larger clock amplitudes are associated with higher noise floors consistent with our result in Fig. \ref{fig:noise_hswing}, which we attribute to charge injection from clock-induced charge \citep{1992ASPC...23....1J}, an effect which becomes a significant component of the measured noise in the photon-counting regime for Skipper CCDs \citep{2022PhRvP..17a4022B}.

\section{Characterization \& Analysis} \label{sec:characterization}
\subsection{Charge transfer efficiency}
Conventionally, charge transfer efficiency (CTE) is described by the transfer efficiency along the parallel register as charge is shifted in columns and along the serial register where charge is passed towards the amplifier. The additional inter-amplifier pixels in the MAS CCD motivates a consideration of the CTE between each amplifier stage. This differs from serial CTE because when charge is passed to the sense node in each amplifier, other charge packets in the inter-amplifier region must be restricted from flowing into neighboring pixels by a sufficiently high potential barrier from the PS gate. We describe both effects in detail by first establishing a baseline serial CTE measured in the first MAS amplifier, then treating the inter-amplifier efficiency between this first stage and subsequent MAS amplifiers. 

\subsubsection{Horizontal register}
\begin{figure}
\centering
\includegraphics[scale=0.39]{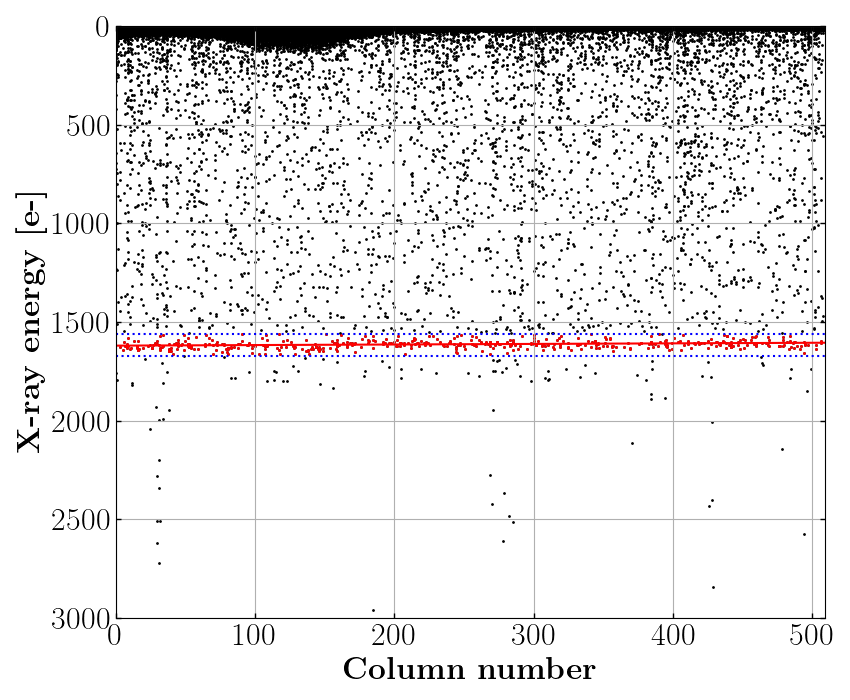}
\caption{Horizontal X-ray transfer does not show evidence of significant charge transfer inefficiency in the serial register before the charge is shifted beyond the first amplifier. A greater charge transfer inefficiency would manifest as a greater tilt in the X-ray line. The tilt of the X-ray line which is the red fit line corresponds to a CTE of $0.999988 \pm 3 \times 10^{-6}$. The fit of the red line is performed for pixels with a total energy within $3\sigma$ of the expected Mn K$\alpha$ line shown by the blue dashed lines.}
\label{fig:xraystacking}
\end{figure}

Horizontal CTE was measured by the standard X-ray transfer method because up until the first amplifier, the CCD is similar in structure to that of a conventional CCD. The CCD was exposed to X-rays from the $^{55}$Fe source for 10 seconds and the gain-corrected signal in each pixel of the output image from the first amplifier was plotted for each column to obtain the stacking plot, as shown in Fig. \ref{fig:xraystacking}, following \citet{2001sccd.book.....J}. The X-ray event line is then fit to within $3\sigma$ around the X-ray Mn K$\alpha$ line, and the tilt of this line corresponded to a CTE of $0.999988 \pm 3 \times 10^{-6}$. This is computed for a stack of X-ray images and with different exposure times as well, with a similar result to within statistical uncertainties; thus, the uncertainty quoted here is the standard deviation of repeating this measurement for several frames. As noted in \citet{2001sccd.book.....J}, this is at the sensitivity limit for X-ray transfer curve CTE measurements, and thus we interpret this to be a lower limit of the CTE in the serial register up until the first amplifier. This demonstrates a traditional horizontal CTE that is consistent with unity and no charge is effectively lost in the pixel charge shifts up to the first amplifier. This is important to establish at the outset because of the distinct effect of inter-amplifier charge transfer inefficiency described in the next section that should be disentangled from any clear CTI upstream from the first amplifier in the MAS output chain.

\subsubsection{Inter-amplifier charge transfer efficiency} \label{actesection}
For a perfect transfer efficiency between the output stages in the MAS serial register, each amplifier should measure the same amount of charge per pixel. Any losses in the extended serial register, which we refer to as inter-amplifier charge transfer efficiency (ACTE), can be quantified by comparing the image output from each amplifier stage.

Charge transfer problems arise when the potential difference between the floating-gate sense node and subsequent horizontal clock phases is too shallow. This new additional gate between the sense node and horizontal clocks is the PS barrier gate. After the charge is read, the PS gate is in the low state for the charge to be transferred to the horizontal phases. If this low PS voltage is too high compared to the floating gate, charge in pixels may not fully exit the sense node. This effect manifests as increasing streaking in signal as charge is shifted down the amplifier chain. In this situation, the output produced in the first output stage is absent of charge transfer problems and streaking; any streaking is observed starting in the second stage with increasing severity as the charge is passed into further stages until the image produced at the last stage is nearly uniformly filled with streaks in each row. This strongly suggests inter-amplifier charge transfer issues because the first amplifier presents no evidence of charge transfer problems even with the same sequence of clocks that leads into each output stage, and instead points to a transfer inefficiency in charge removal from and shifting between output stages.

\begin{figure}
\centering
\includegraphics[scale=0.4]{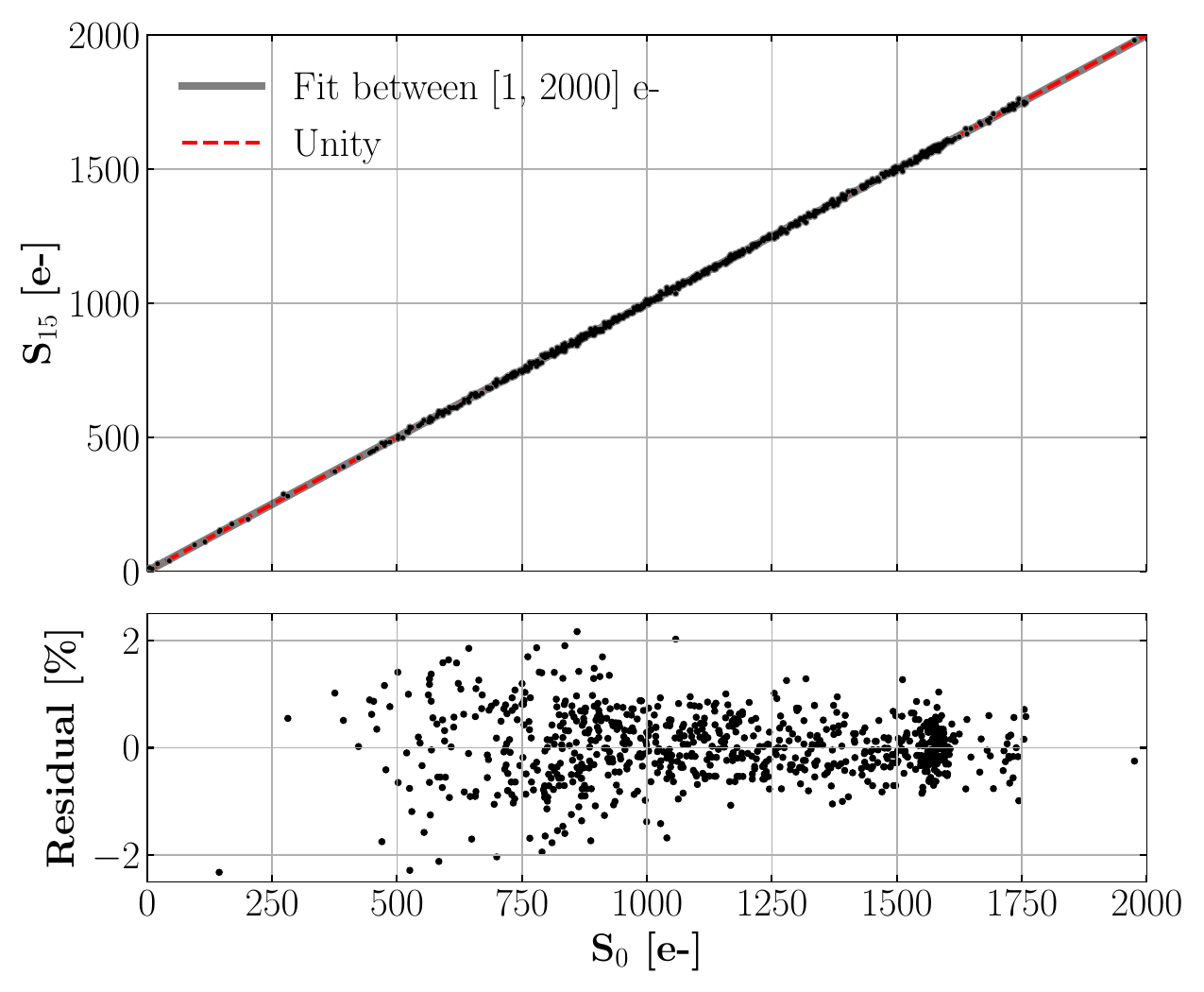}
\caption{Peak X-ray signal (in electrons) in a pixel in amplifier 1, $S_{1}$, is compared with the same pixel in amplifier 15, $S_{15}$ for an example image. For perfect inter-amplifier charge transfer efficiency, the slope should be unity (red dashed curve). Each X-ray hit is plotted and linearly fit (gray curve) to measure any deviation from unity. The cluster of signal charge near 1600 e- arises from the X-ray events around the Mn K$\alpha$ line. Fitting with the high scatter in this method is primarily what limits the precision of this measurement.}
\label{fig:ctefitting}
\end{figure}

To measure the low-signal ($\sim$$10^3$ electrons) inter-amplifier CTE (ACTE), we expose the CCD to X-rays. After overscan correction, the location of each X-ray event is identified by a standard source extraction algorithm \citep{Barbary2016}, where aperture photometry is performed to obtain the conversion gain. \new{This process for this was described in $\S$\ref{gaincalibration}.} The gain is computed for each MAS amplifier and each image output was converted to electrons from ADU to compensate for any gain variations between amplifiers. The peak pixel value in electrons and its corresponding coordinate for a cluster of pixels affected by an X-ray photon in the first MAS amplifier is compared against the number of electrons at this same pixel location in subsequent serial amplifiers. If charge transfer efficiency is perfect, the pixels should have an identical number of electrons at each MAS output stage. As Fig. \ref{fig:ctefitting} demonstrates, deviations from a unity slope is indicative of imperfect charge transfer between amplifiers. After correcting for the respective gains of the amplifiers, we find that the inter-amplifier efficiency between the first and last amplifiers is $0.9984 \pm 0.0033$ below the 2000 electron signal level, the energy range where most of the X-ray induced charges are concentrated. This was computed from measuring several X-ray exposures and averaging the result; the relatively high scatter in the measured ACTE value when repeating this method for many images drives the elevated uncertainty, motivating us to pursue a complementary approach to more precisely quantify the ACTE effect.

\begin{figure}
\centering
\includegraphics[scale=0.4]{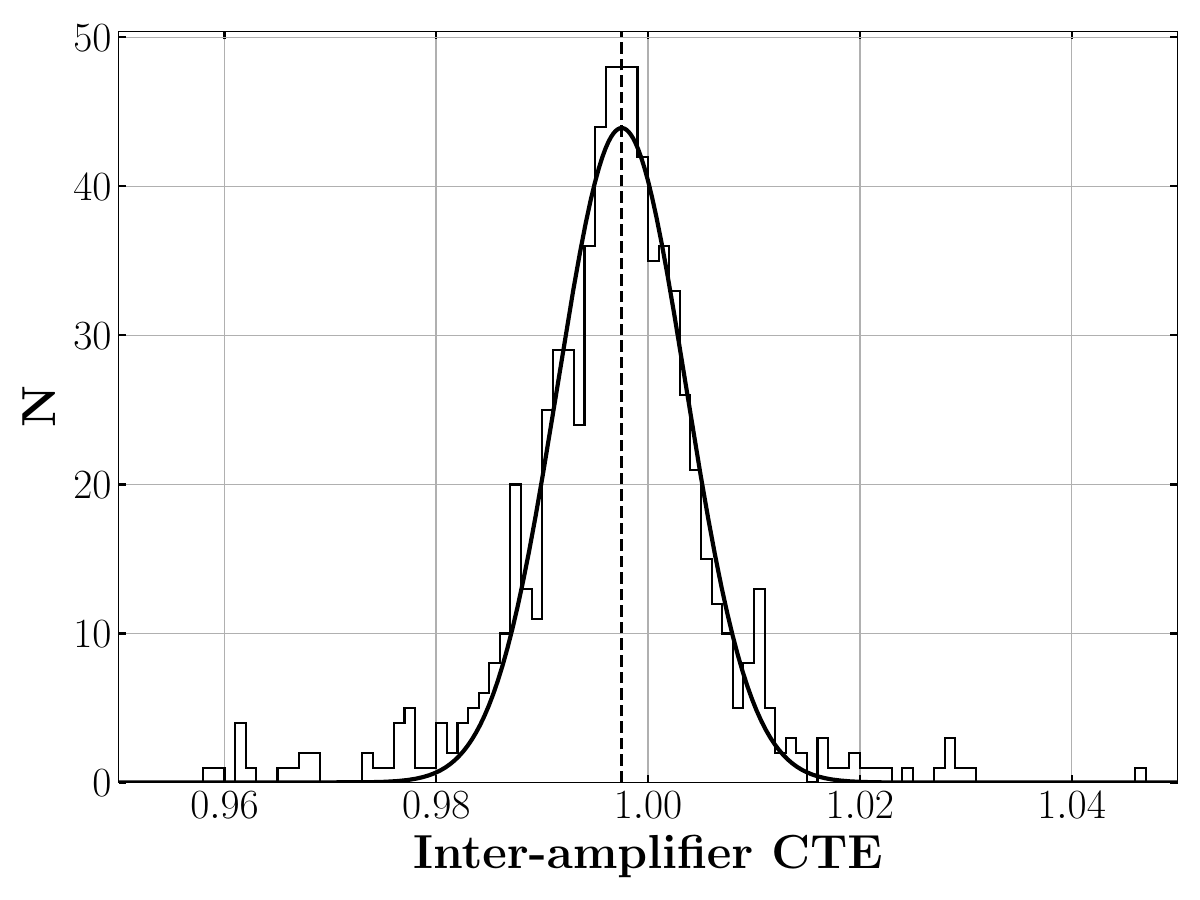}
\caption{Inter-amplifier charge transfer inefficiency is calculated by modeling the trailing pixel to the pixel with the peak X-ray signal in an X-ray hit following Eq. \ref{eq:ctemodel}. Accumulating statistics for all X-ray hits in an exposure gives a measurement of the inter-amplifier charge transfer efficiency at the mean of the Gaussian fit, which as shown in this plot is 99.8\% across all 16 output stages, comparing the first and the last amplifier.}
\label{fig:ctestats}
\end{figure}

This secondary method we have developed to quantify the ACTE involves linearly modeling charge transfer efficiency between the peak X-ray flux pixel in the first output stage with the adjacent pixel in the direction opposite to the serial shifts. This simple model, parameterized by the charge transfer inefficiency, is described by
\begin{multline} \label{eq:ctemodel}
S_M (x_p + 1, y_p) = (1-\text{CTI}) \cdot S_0(x_p+1, y_p) \\ + \text{CTI} \cdot S_0 (x_p, y_p)
\end{multline}
where $S_M(x,y)$ is the signal, in number of electrons, at pixel coordinate $(x,y)$ of amplifier $M$, and $(x_p, y_p)$ denotes the pixel with the highest flux from a given X-ray hit. In this model, the number of electrons in the pixel next to the X-ray peak pixel opposite to the serial shift direction (i.e., trailing pixel) is expressed as $S_M(x_p+1, y_p)$. This quantity is the sum of two components to first order: the number of electrons in this pixel as measured in the first amplifier (i.e., amplifier 0) with some loss due to non-zero CTI and the number of electrons lost from the peak X-ray pixel in the first amplifier due to non-zero CTI. A CTI value can be calculated for each X-ray hit on the image, and the distribution of these values can be fitted with a Gaussian to statistically estimate the ACTE as exhibited in Fig. \ref{fig:ctestats}. Using our lowest noise voltage configuration, we find that our lower limit for the inter-amplifier charge transfer efficiency between the first and last MAS amplifiers is $0.9973 \pm 0.0001$ by this statistical approach. This is in agreement with the simpler first approach illustrated by Fig. \ref{fig:ctefitting} to within uncertainties. \new{Since this is measured at the level of 0.3\%, the effect of adding additional higher order terms is negligible and would not change our results.}

\begin{figure*}
\centering
\includegraphics[scale=0.4]{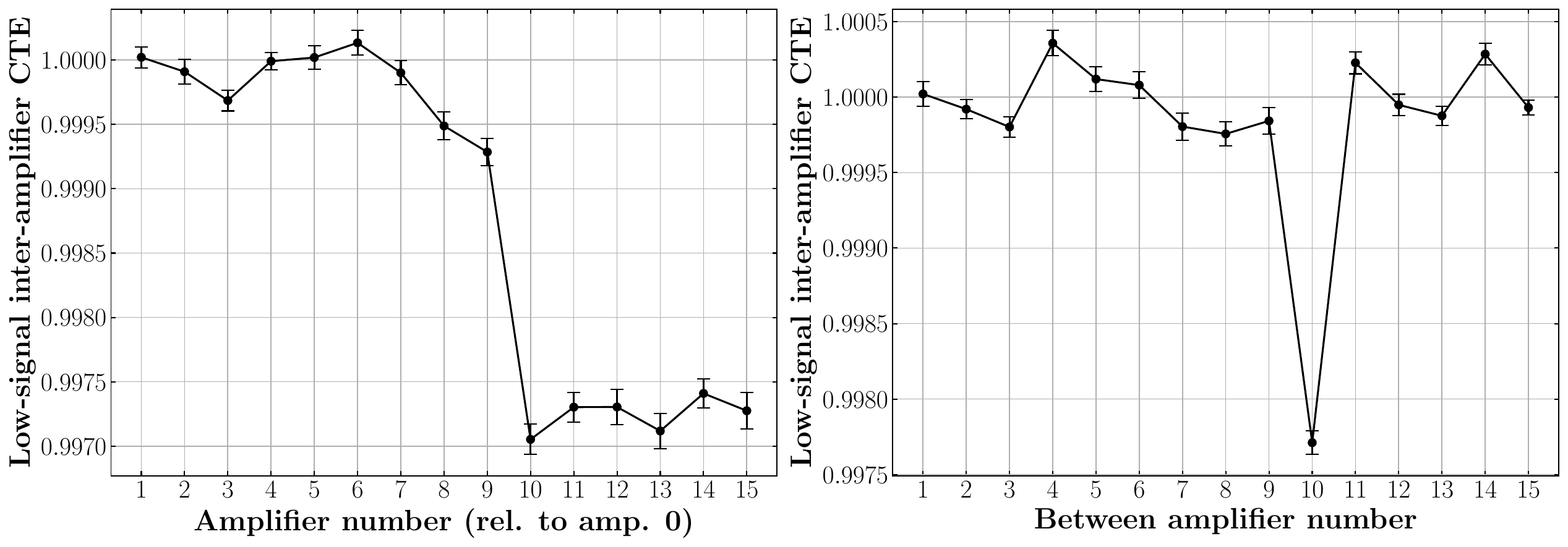}
\caption{X-ray measurements of inter-amplifier charge transfer efficiency across successive output stages based on CTI modeling. The left panel shows the CTE at each amplifier relative to the first amplifier (amplifier 0). For example, at amplifier 8, the output between amplifier 8 and amplifier 0 is being measured. The right panel shows the inter-amplifier charge transfer efficiency between each amplifier. For example, at amplifier 8, the transfer efficiency between amplifier 7 and 8 is being measured.}
\label{fig:ctecomparison}
\end{figure*}

The comparison between the transfer efficiency of each consecutive amplifier with the same measurement shown in the right panel of Fig. \ref{fig:ctecomparison} reveals that amplifier 10 stage has about a $0.2\%$ lower than expected signal measured, suggesting a possible device-specific issue such as a localized charge trap from lattice defects in the register between amplifier 9 and 10. In the next transfer from amplifier 10 to 11, we observe a recovery in the ACTE, meaning that charge was properly transferred to amplifier stage 11 from 10. However, despite the lack of measured CTI between each output stage from amplifier 10 onwards, the measured charge in these latter stages were all below the expected signal level read out by amplifier 0. This is shown by the drop in the ACTE curve relative to amplifier 0 at amplifier 10 where this efficiency is at a plateau around $0.9973$. Since charge in each pixel must pass through each output stage, losses due to defects along the way can become cumulative and can be a potential drawback for the MAS design. While it may be possible to reduce the impact of obstacles in the channel that degrade charge transfer performance by tuning clock voltages and timing, this first prototype device and existing CCD controllers that we are aware of do not support individual clock lines to multiple output stages and are impractical for scaling to larger amplifier count devices. Excluding the possible defect in our detector between amplifier 9 and 10, we observe promising ACTE $>0.9999 \pm 0.0001$ at the level of up to $2 \times 10^3$ e-.

However, X-rays are limited in energy range to the absorption lines of the $^{55}$Fe source, so probing for transfer problems at higher signal levels at the $10^4$ level and above requires an alternate approach. To reach higher illumination levels, the LED projector was triggered \new{by specifying an exposure time during which the LED would be on and the CCD would be in expose mode. After the exposure time elapsed, the LED would turn off and the CCD readout would begin. By synchronizing the exposure time with the LED uptime, the LED acts as a virtual shutter with no mechanical shutter needed in our setup.} The LED was connected externally to an external controller that connected to the synchronization lines of the Hydra and included options for adjusting the LED intensity.

Our first approach is based loosely on the extended pixel edge response (EPER) method, which is a standard horizontal CTI measurement technique for conventional CCDs relying on flat field images to determine the amount of deferred charge in the overscan region \citep{1988SPIE..982...70J}. However, in addition to charge transfer in the inter-amplifier pixels, since charge enters and is removed from the charge sensing areas multiple times in the MAS architecture, the traditional EPER formulation is inadequate for analyzing the potential charge transfer loss and deferral between one output stage and the next. Thus, we follow Eq. \ref{eq:ctemodel} to compute the inter-amplifier CTI given the signal measured in the first overscan column between the first and last amplifiers. Applying this technique with EPER-like data exploits the sharp edge between the active and overscan regions. Because the overscan region is not expected to contain incident signal if the CTI is ideally zero, the sharp edge provides a sensitive probe of charge in the overscan area after shifts through multiple sense nodes that is independent of the surface uniformity of the illumination from imperfect optics.

\begin{figure}
\centering
\includegraphics[scale=0.38]{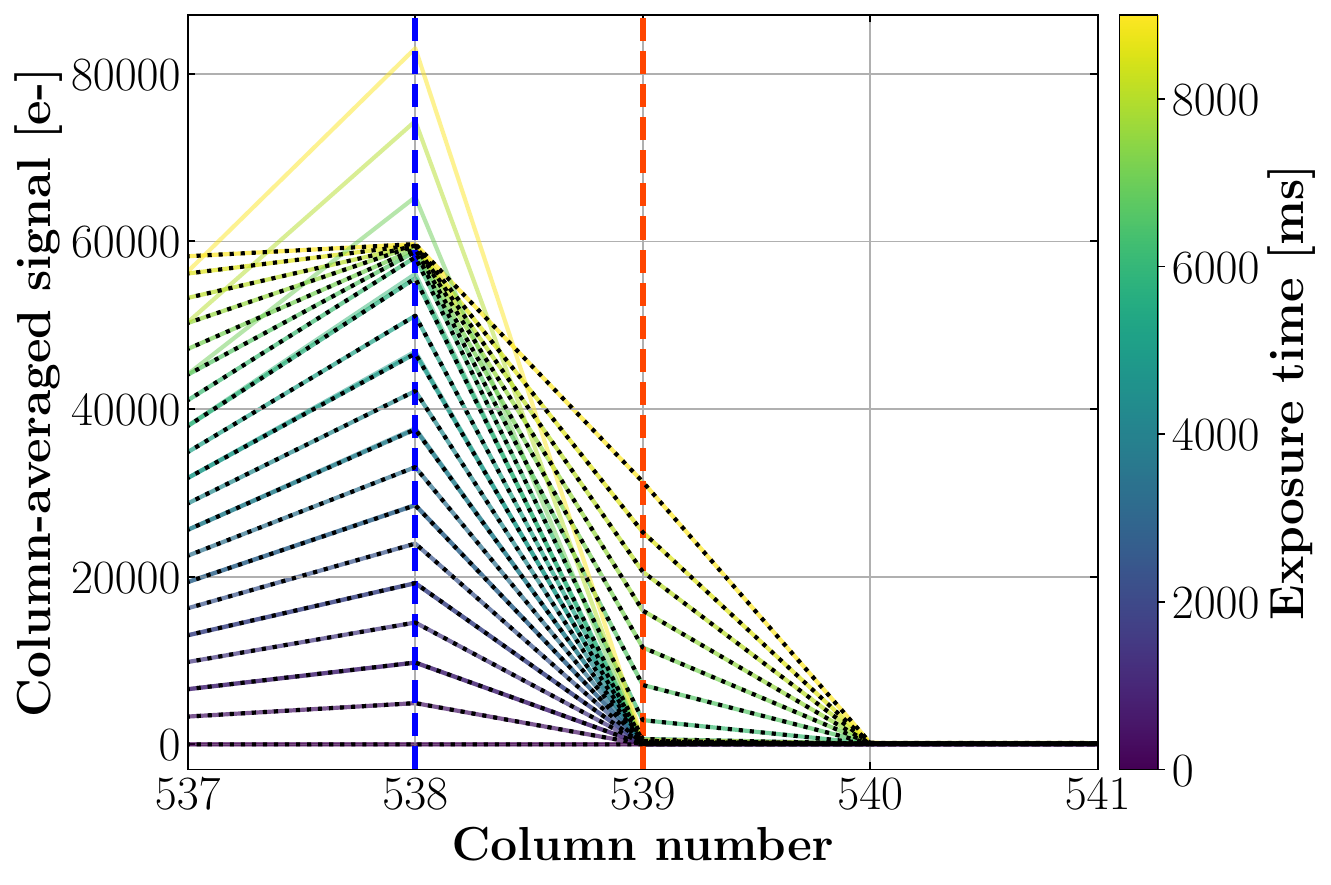}
\caption{Horizontal profiles around the boundary of the active and overscan regions at varying exposure times. The solid lines are profiles from the first amplifier (amplifier 0) and lines with dots are profiles from the last (amplifier 15) with the color spectrum representing the exposure time. The blue vertical dashed line indicates the last column in the active area and the orange dashed line the first overscan column. Blooming at the edge of the active region into the overscan is caused by the imperfect charge transfer between amplifiers as shown by the degradation of the profiles at lower signal levels than the apparent full well in amplifier 15. \new{For even longer exposure times ($\gtrsim$ 10 s), we see the profile of the first amplifier bloom into the overscan due to the over saturation of the pixels in the active area (these profiles are not shown in this plot to maintain clarity).}}
\label{fig:horizontalprofiles}
\end{figure}

Illumination at varying signal levels was obtained by increasing the exposure time in increments of 250 ms up to 28 s in a linear ramp, with exposures taken at each exposure time. Using a reduced ADC gain, this ramp extends up to $7.8 \times 10^4$ e- in incident signal level. The gain was again calibrated at this ADC configuration for each amplifier and all images were bias-subtracted with a master bias for each amplifier. \new{The master bias was obtained by averaging a stack of 40 bias frames on a per-pixel basis.} The horizontal profiles are shown in Fig. \ref{fig:horizontalprofiles}, where the signal average of each column of pixels up to the vertical overscan are compared at various exposure times. The last active column is 538 and like typical LBNL $p$-channel CCDs contains more charge than other pixels in the flat field. This effect may arise because charge from the flat field diffuses from material outside of the array at the edges \citep{1989ITNS...36..572J} and because fringing fields result in pixel size variations at the edges \citep{2014JInst...9C4001P}. \new{If we consider the first amplifier, as the signal level increases with the exposure time and saturation approaches ($>8 \times 10^4$ e-) the charge begins overflowing into the overscan region. As the pixels in the active area completely saturate, the horizontal profile over the width of the CCD develops a characteristic flat top.} This is an expected effect present in each amplifier that reflects the physical pixel full well given our operating parameters. However, the horizontal profile for amplifier 15 shows that the charge overflow into the overscan pixels occurs earlier compared to the first amplifier, pointing to an inter-amplifier CTI effect that dominates the overall charge losses along the extended MAS register even before the pixel full well capacity is reached.

\begin{figure}
\centering
\includegraphics[scale=0.28]{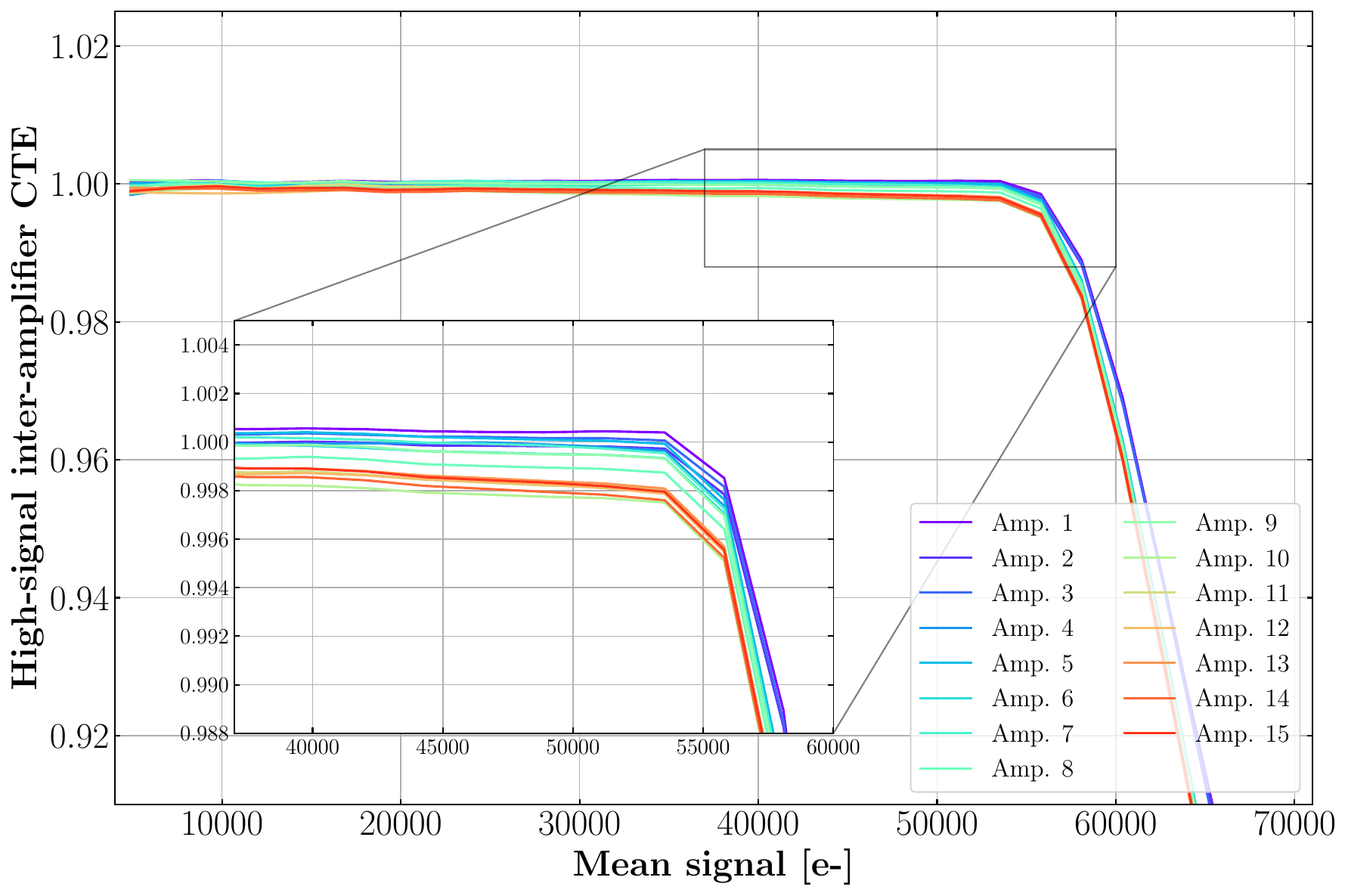}
\caption{Inter-amplifier CTE at high signal levels shown for each amplifier stage relative to amplifier 0 show a catastrophic inter-amplifier CTE decline past 55,000 e-. The inset shows in detail this drop off at high signal.}
\label{fig:flatscte}
\end{figure}

To diagnose the amount of charge where the inter-amplifier charge transfer efficiency breaks down, we apply Eq. \ref{eq:ctemodel} for the mean of the final active column as the ``peak pixel" and the first overscan column as its ``trailing pixel." This is evaluated at the first amplifier stage in relation to each of the subsequent stages from 5,000 to 80,000 e-. The results for this ACTE analysis is summarized in Fig. \ref{fig:flatscte}, where the measured ACTE for each output stage relative to amplifier 0 is plotted, showing the signal regime at which the amplifier-to-amplifier transfer drops. The inter-amplifier CTI between amplifier 0 and 1 is consistent with unity until above 53,000 e- where ACTE degrades swiftly. However for amplifier 15, the ACTE has a slightly steeper slope as seen in the inset of Fig. \ref{fig:flatscte} and rapidly drops starting around 50,000 e-. This initial decay leads to a rapid turnover starting just under 55,000 e- where charge transfer between amplifiers becomes unreliable with increasing blooming of charge into the overscan columns. For scientific requirements where conventional CTI $< 10^{-5}$, this imposes a strict upper limit on the effective dynamic range of this MAS device with our operating parameters at $50,000$ e- for reading out through all 16 channels. This demonstrates that the amplifier-to-amplifier transfer efficiency breaks down well before the actual physical well depth of the pixels given our clock voltages, and thus acts as an ``effective full well." This behavior where the ACTE rather than the physical pixel well depth imposes an effective full well is analogous to how the output-amplifier voltage swing limits the full well of conventional CCDs \citep{2017JInst..12C4018B}. \new{We note that although the inter-amplifier CTI at amplifier 10 is larger than $10^{-5}$ in Fig. \ref{fig:ctecomparison}, the requirements for pixel-to-pixel CTE are different from the inter-amplifier CTE. This is because the nominal $10^{-5}$ CTI requirement is an average over thousands of pixel shifts, while the ACTE is measured over 16 amplifiers and the maximum combined effect is a 0.3\% inefficiency. For a $10^{-5}$ CTI standard over 1000 shifts, this translates to an integrated CTI requirement of $< 1\%$, and this is met even with the amplifier 10 defect.}

Although $\sim$50,000 e- dynamic range is limiting for imaging which typically requires $\gtrsim 90,000$ e- full well minimum \citep[e.g.,][]{2009JInst...4.3002R,2015AJ....150..150F}, this is sufficient for spectroscopic surveys such as DESI and photon-starved applications. In exploring additional voltage configurations, we find that it is possible to change the signal level at which the drop off of the ACTE occurs through different combinations of the H1 and PS clock swings. The critical step in the readout sequence of clock states where PS and H1 have direct impact is state (c) in Fig. \ref{fig:transferdiagram}, immediately following the measurement of charge in the sense node. For example, smaller voltage swings for H1 where its lower state is raised above -12 V, maintaining a potential difference ranging between 1 V to 5 V between the low voltages of PS and H1, moves this ACTE drop off to even lower signal levels. With high signal levels and a collecting potential phase that is too shallow, when the charge is transferred into the inter-amplifier pixels after measurement at one output stage, not all charge in the sense node is transferred and unwanted charge diffusion back into the sense node before the PS gate is raised as a potential barrier may occur, increasing the charge loss down the serial line of amplifiers. These observations suggest that larger H1 clock swings or a higher PS low state at least at the clock amplitudes of the vertical clocks (which determine well depth in the array) would potentially enable a greater dynamic range than what we are able to measure, given the constraints of the $\pm 12$ V range of the Hydra electronics. Increasing clock amplitudes to increase the ability to transfer charge at high signal levels must be tempered with the effective read noise trade-off, since greater horizontal clock voltage swings as presented in Fig. \ref{fig:noise_hswing} correlate with higher read noise $> 1.5$ e- rms/pix. Therefore, key elements such as the target noise and effective full well required must be well-defined and optimized on simultaneously because the multivariate relationships between the clock voltages and the resulting behavior are interconnected in the MAS architecture, as single-parameter only optimization is likely to jeopardize another aspect of device performance.

Another method explored for this analysis involved projecting a pattern of a grid onto the detector surface. At the front of the LED, a mask lens with a grid pattern was inserted and manually aligned to the CCD rows and columns. \new{Alignment was done by rotating the LED projector with the mask lens attached, taking a sample image, and evaluating whether the center lines of the illuminated grid columns and/or rows had flat profiles, since in the illuminated grid regions the signal level should be flat.} The grid was focused by minimizing the width of the illuminated column that the LED projects onto the active area. The grid pattern mask ensures that the illumination level is consistent in a given column or row of the grid. As with flat fields, different signal levels in the illuminated areas of the grid could be obtained by varying the exposure time. To measure the CTE through the full dynamic range up to the full well capacity, we took images in a ramp of increasing exposure times up to 30 s at a low gain ADC setting.

\begin{figure}
\centering
\includegraphics[scale=0.44]{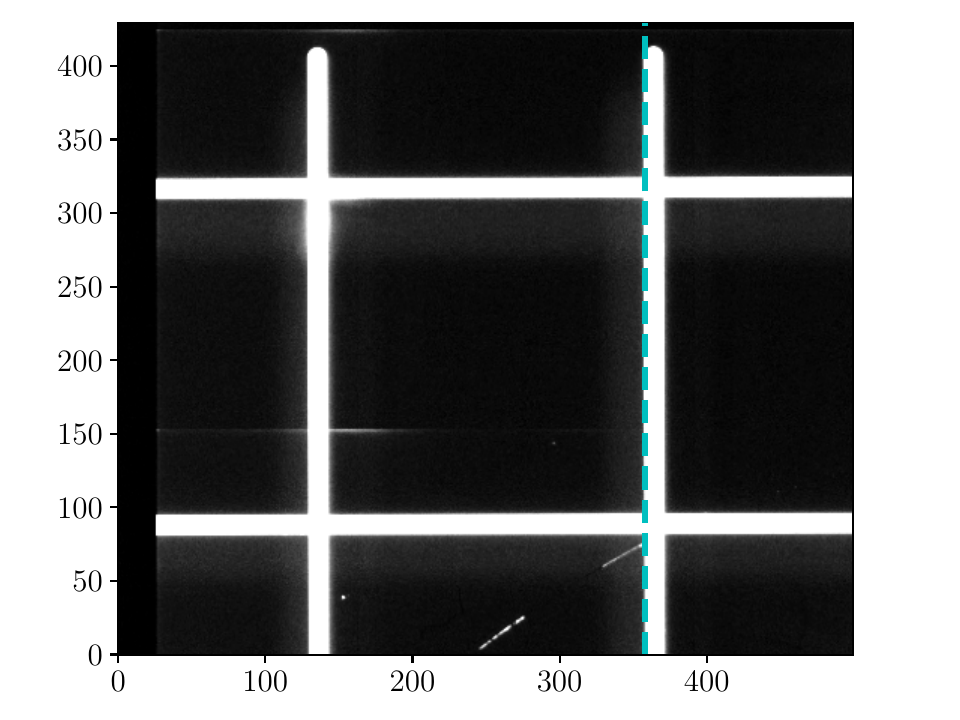}
\caption{Grid mask projection onto the active area of the CCD with the edge of the illuminated strip where profiles are taken denoted by the dashed line. This dashed line in the same location is also denoted on the profiles shown in Fig. \ref{fig:gridbadcte}. Stray light from a few isolated micron-scale scratches are visible in the dark areas.}
\label{fig:gridimage}
\end{figure}

\begin{figure}
\centering
\includegraphics[scale=0.4]{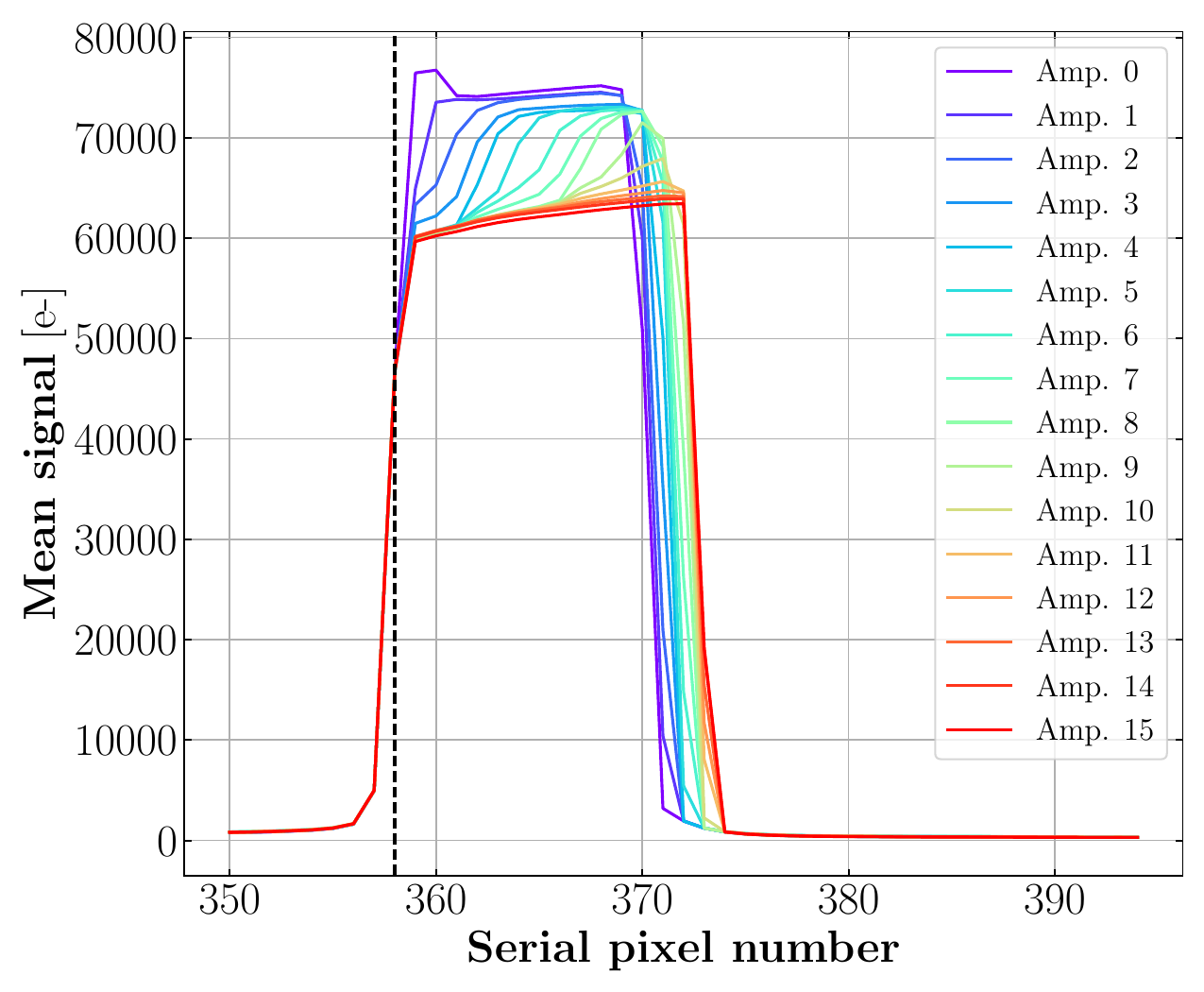}
\caption{Horizontal profiles around the illuminated grid columns plotted for each amplifier demonstrates significant inter-amplifier charge transfer issues that compounds with every successive output stage. The vertical dashed line denotes where there is a ``cliff" along the border between the illuminated and non-illuminated columns. \new{This is the corresponding dashed line shown on the image in Fig. \ref{fig:gridimage}.} The signal flux at this cliff is used to evaluate the ACTE.}
\label{fig:gridbadcte}
\end{figure}

To account for non-uniformities in our mask, vertical profiles were measured for the illuminated grid areas. We base our measurements on areas of the grid that do not intersect, because pixels in the intersection of illuminated grid areas have higher signals that have contributions from optical effects that are challenging to disentangle from the potential imperfect charge transfer that we are measuring. To quantify CTE with this setup, we measured the averaged horizontal profiles of an illuminated column segment of the grid for each amplifier. This column was carefully chosen where the vertical profile of the segment is flat and is at the leading edge of the illuminated strip, illustrated in Fig. \ref{fig:gridimage}. The leading edge between dark and illuminated pixels is ideal since a charge transfer issue is expected to result in deferred charge starting at the illuminated region which blooms past the trailing edge. For perfect CTE, the profiles for each output amplifier stage should be identical and completely overlap, whereas charge transfer issues result in profile differentiation illustrated in Fig. \ref{fig:gridbadcte}. Significant CTI at high signal levels can be observed on images as a broadening of the illuminated grid column on the trailing pixel side.

\begin{figure}
\centering
\includegraphics[scale=0.34]{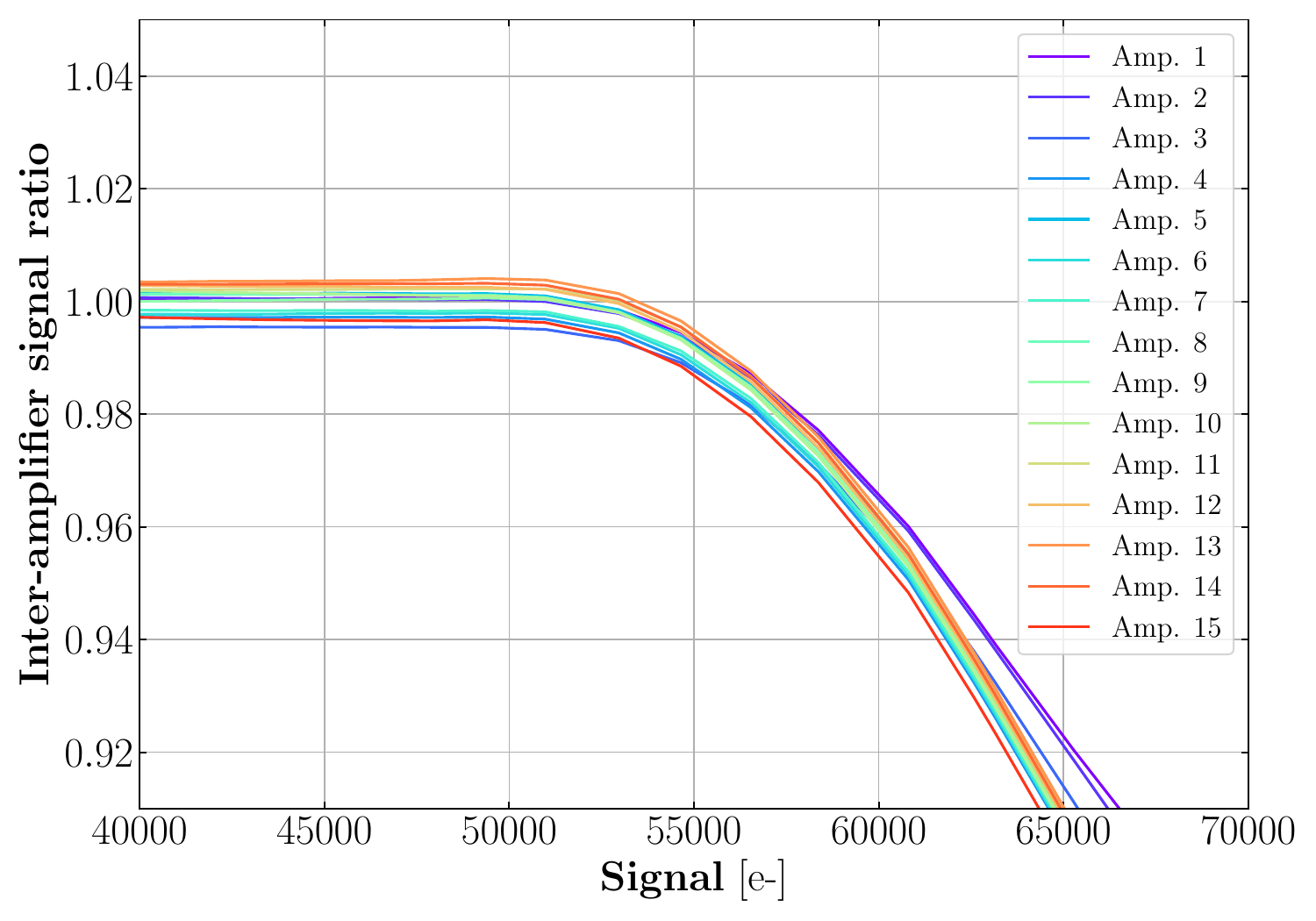}
\caption{The ratio of the mean signal between the first and subsequent amplifiers as a function of the signal level. The column number where the mean signal is evaluated is denoted by the dashed vertical line shown in Fig. \ref{fig:gridbadcte}. The degradation of ACTE shown here is consistent with what was found with Fig. \ref{fig:flatscte}.}
\label{fig:gridcte}
\end{figure}

The results using the grid illuminated at varying signal levels is presented in Fig. \ref{fig:gridcte}, where the ratio of the signal in each amplifier at the location of the column along the edge of the grid is interpreted as a measure of the ACTE. Although using this technique gives noisier results than with our EPER-like measurement, the shape and drop off of the curve is nearly identical to what we find in Fig. \ref{fig:flatscte}. Inter-amplifier charge transfer becomes unsustainable and drops rapidly at signals between 54,000--55,000 e-. More precise measurements are limited for this approach by imperfections in the alignment of the grid to the pixel columns, micron-scale scratches on the surface of the mask, and possible non-uniformities in the optics and the mask itself that are not straightforward to resolve. These are extraneous effects at the $0.8\%$ level that dominate over $< 0.1\%$ effects from ACTE that play a role prior to the complete drop off of the ACTE.

\subsection{Full well capacity} \label{sec:fullwell}
While low-signal spectroscopy typically does not require high full well capacity, maximizing full well elevates the versatility of a detector by increasing its dynamic range. This is an important consideration for spectroscopic observations capable of capturing the full intensity of bright lines and photometrically imaging complete transient event light curves from the low-signal tails of the rise to peak flux without flat-topping due to saturation. For amplifier 0, we find that the full well capacity at a low ADC gain setting with our low-noise voltages is $\sim 80,000$ e-, as observed with the amplifier 0 horizontal profiles in Fig. \ref{fig:horizontalprofiles} at the highest signal levels. However, this full well is only valid for amplifier 0 because this level cannot be reached due to ACTE degradation. Full well measurements for multiple amplifiers must appropriately account for the inter-amplifier charge transfer efficiency, which adds excess charge that has bloomed over in the serial direction at very high signal levels.

Our measurements with the inter-amplifier charge transfer efficiency indicate that there is an ``effective full well" in the MAS architecture bounded by non-linearities at high signal that is distinct from the physical well depth of the pixel given the clock voltages. These non-linearities result from the highly inefficient charge transfers between amplifier sense nodes in the MAS register that dominate all other effects starting at the 50,000 e- signal level. This is specifically the limit contributed by amplifier 15, the final output stage, and thus can be interpreted as an effective charge level constraint for the device. Because this effective full well is not set by the pixel well depth but by the PS-H1 clock voltages that dictate ACTE, it is another parameter that should be optimized on, akin to how the sense-node conversion gain impacts the full well of conventional CCDs \citep[e.g.,][]{6742594, 2017JInst..12C4018B}.

\subsection{Linearity} \label{sec:linearity}
Detector nonlinearity was explicitly measured by acquiring flat field images in a sequence of exposure times that provide different illumination levels. Since our system was not set up for a mechanical shutter and because shutters also have intrinsic limitations on precision timing, the LED illumination source was kept on at fixed intensity even between exposure readouts to ensure that the timing of the LED warm-up and on-off cycle was not a systematic uncertainty in our measurement. In idle mode, the Hydra continuously clocks the parallel register until an exposure is triggered for a time, where the clocks are frozen for charge to collect in the active area. Without a mechanical shutter with the LED constantly on, charge will accumulate in the array that will compound into the subsequent exposure if there is insufficient idle time for the charge collected during the readout time to be clocked away. To minimize timing variation and obtain as flat a background as possible in this setup, an idle duration of 20 s was inserted between each exposure, and a zero-second exposure was interleaved between each image acquired with nonzero exposure time to correct for the baseline signal on the detector at any given time by the fixed LED light. Thus, bias correction was accomplished by subtracting each nonzero exposure time image from its preceding zero frame. Three exposure pairs per unique exposure time were taken with an exposure time sequence that was in randomized order to further reduce systematic effects.

\begin{figure}
\centering
\includegraphics[scale=0.38]{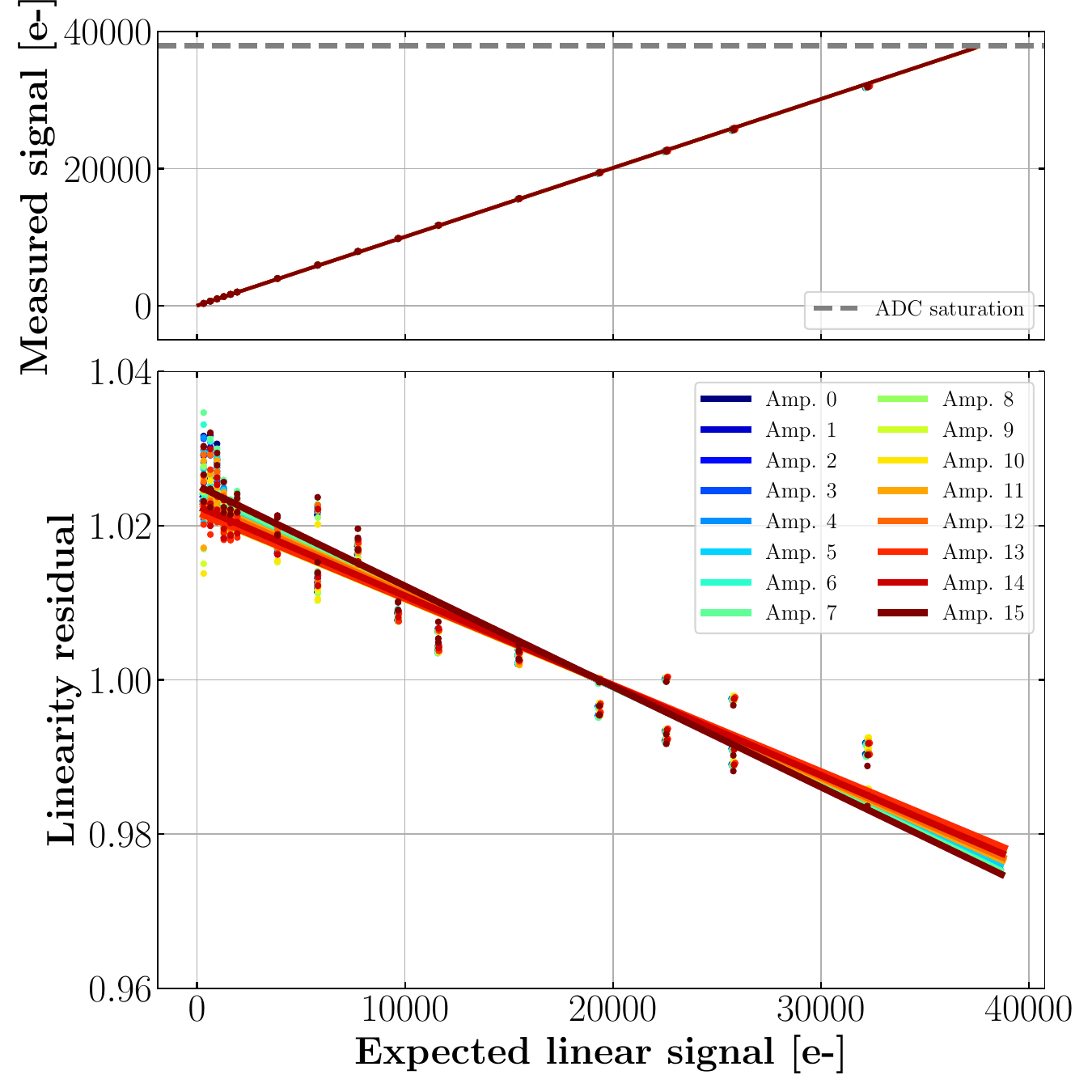}
\caption{Nonlinearity of all amplifiers for the MAS CCD, with data points shown in the upper panel and a linear curve fit in a single color assigned to each amplifier. The dashed horizontal line indicates the signal at which the ADC numerically saturates, since we selected a high gain setting to acquire these exposures. The bottom panel quantifies the residual as a function of the expected flux for the optimal fit linear model. \new{The gains were calibrated with X-rays, as described in \S\ref{gaincalibration}.}}
\label{fig:ampallnonlinear}
\end{figure}

We acquire data across a broad dynamic range of 28 illumination levels, ranging from 330 e- to about 35,000 e-, corresponding to where the ADC saturates and before where the inter-amplifier charge transfer efficiency starts a gradual decline. This allows us to capture detector nonlinearity in a regime that is not dominated by ACTE effects. The data is shown on the upper panel of Fig. \ref{fig:ampallnonlinear} for each output stage. After bias correction, the average signal of the image was computed by taking the median value of a clean, central subregion of the active area. First degree polynomial fitting by minimizing squared errors was performed on this measured signal flux up until 35,000 e-. This model was used to compute the expected signal level plotted as the independent variable in Fig. \ref{fig:ampallnonlinear}, resulting in the linear curves shown on the upper panel. To quantify the linearity, the residuals to this linear optimal fit are calculated in the bottom panel, where the residual quantity is defined as the ratio of the measured flux level and the flux expected from the linear model, and is proportional to the exposure time. A perfectly linear response would yield a flat line at unity. This linear model demonstrates that for each amplifier there is a consistent $\pm 2.5\%$ nonlinearity in the MAS CCD across the 330 e- to 35,000 e- dynamic range. This mild nonlinearity becomes more significant as the signal nears the effective full well with the onset of inter-amplifier charge transfer efficiency deterioration, and thus is excluded from the linear fitting. These results demonstrate that the MAS CCD is on par with or slightly better than typical DECam CCD nonlinearity characterized using a similar approach \citep{2017PASP..129k4502B}. At the very low signal levels under a few hundred electrons, this method is limited by systematics that may vary with time such as light leakage, and thus we do not probe signals at this very low regime. However, we note that it is possible to probe very low signal levels by operating MAS CCD in photon-counting mode as a Skipper CCD using repeated skipping sampling to count the number of electrons per pixel in relation to the number of ADUs output by the ADC. This is essentially an absolute measure of gain stability and was studied for Skipper CCDs by counting electron peaks up to 1800 e- where a $< 2.5\%$ nonlinearity was found at this low photon-counting regime \citep{2021NIMPA101065511R}. Our new readout electronics being developed will enable the capability to switch seamlessly to the photon counting mode, where we will be able to extend our measurements down to a few tens of electrons. An initial demonstration of photon-counting with a MAS CCD by ``skipping" has been shown in \citet{10521851}, using a low-threshold acquisition (LTA) controller optimized for dark matter experiments.

\subsection{Crosstalk}
Multiple channels measure charge simultaneously in the MAS CCD, introducing an elevated risk of electronic crosstalk between channels that generates ghost images that can contaminate neighboring channels. One indication of crosstalk we observed especially in flat fields was the slightly different baseline levels within the overscan and prescan regions of the image. In column-averaged horizontal profiles of the prescan and overscan regions, after the sharp drop-off in signal from the active region, the first 16 pixels into the overscan form a ``step" before dropping to a stable baseline. This 16-pixel wide step corresponds to the number of inter-amplifier pixels separating the output stages, and is characteristic of crosstalk generated by amplifiers that are reading pixels at the same time for pixels physically ahead and behind horizontally. The signal difference between this 16-pixel step and the overscan baseline scales linearly with the signal in the final active column until inter-amplifier CTI losses cause blooming into the overscan itself. Using this feature to estimate the level of this ghosting signature present in each channel, we found the upper limit of this crosstalk to be $0.08\%$ of the signal in the final active column in the amplifiers 4-8, in the middle of the chain. The amplifiers with lower levels of ghosting were 0-3 and 9-14, where the crosstalk amplitude was measured to be $<0.04\%$, comparable to common requirements for multi-output conventional CCDs \citep[e.g.,][]{2001ExA....12..147F,2015JInst..10C5010O}.

\section{Upcoming plans} \label{sec:plans}
For photon-limited observations required for the more aggressive space imaging application, single-photon resolution is essential and necessitates a combination of further reductions in noise per amplifier, additional MAS output stages, and taking advantage of the ability to ``skip" with the floating-gate amplifier. In the photon-counting regime, systematic noise backgrounds normally hidden underneath read noise in silicon detectors become dominant, imposing a potential detection limit that must be characterized. When the magnitudes of the carriers created from thermal leakage (dark) current and clock-induced charge are comparable to incident optical signals, the photons are to be detected become indistinguishable from the background generated from the detector itself. While amplifier thermal leakage is typically managed by cooling the detector, clock-induced charge spuriously generated during readout is a well-documented but not yet fully-understood single-electron event mechanism that single-photon resolving Skipper and electron-multiplying CCDs are sensitive to detecting \citep[e.g.,][]{2001sccd.book.....J, 2022PhRvP..17a4022B, 2014SPIE.9154E..0CW}. Because the MAS CCD has additional clocks owing from inter-amplifier pixels in its extended output register, investigating the nature of spurious charge in the photon-counting regime is especially important, and our next phase of testing enables output gate clocking for ``skipping" capability to extensively quantify its dependence on clock amplitude, edge shape, and pixel rate. Re-sampling charge to reach photon-counting resolution also provides the additional benefit of direct gain self-calibration and very low-signal linearity measurements during potential daytime detector calibrations in survey operations. New readout schemes that benefit from multiple amplifiers will also be studied. To accommodate these functionalities with a large format MAS CCD, a 64-channel readout electronics system is being designed that will be used to carry out these tests.

Thinned MAS CCDs with 8 and 16 channels have been fabricated with the same high-efficiency ITO/ZrO$_2$ AR coating applied as DESI detectors \citep{2017JAP...122e5301G}. Tests are underway to analyze the optical properties of these devices. Potential future designs being considered such as four-corner MAS amplifier layout with 32 channels each to accommodate simultaneous readout from quadrants is anticipated to reach sub-electron read noise and may be suitable for space-based imaging and spectroscopy applications.

\section{Conclusions} \label{sec:summary}
We have packaged and instrumented a thick MAS CCD prototype designed at LBNL using a synchronized adaptation of the DESI readout electronics with all 16 channels fully functional. The development of the MAS CCD was driven by the motivation to reduce the readout time required by Skipper CCDs to achieve single to sub-electron read noise by placing multiple video output stages along a longer serial register where pixel charge can be measured multiple times. Reducing the readout time while maintaining the capability to sample charge multiple times to lower read noise is critical for enabling next generation ground-based astronomical spectroscopy and especially for all-sky surveys where exposure time is limited by cosmic ray contamination and practical observational strategy constraints. Future spectroscopic surveys including a DESI upgrade or a Spec-S5 will require a single-electron read noise floor to reach fainter objects at higher redshift, particularly in blue optical wavelengths were detector read noise currently dominates the noise budget. Multiple on-chip amplifiers also provide additional layers of redundancy against amplifier failures with the ability for charge to non-destructively pass through dead or high-noise amplifiers which regularly plague CCDs. Since the information carried by the charge is preserved, only functioning amplifiers with acceptable noise can be selected for readout or for optimizing the compound read noise. All of these features would benefit any survey especially when detectors cannot be replaced during operations.

After voltage optimization on noise, we have demonstrated a near single-electron compound read noise of 1.03 e- rms/pix with a single read per amplifier at a pixel time of 26 $\mu$s/pix. This is nearly a factor of 8.5 less in readout time per pixel compared to that of a Skipper CCD performing at an equivalent noise level \citep[e.g.,][]{2024PASP..136d5001V}. With this performance demonstrated as a prototype device, a MAS CCD with twice as many amplifiers would be a promising contender for a detector upgrade in the second phase of DESI and as a candidate detector for a Spec-S5 spectrograph. These optimization tests have shown the interplay amongst the noise, full well, and pixel charge transfer efficiency between the extended multiple stage outputs of this architecture. To investigate the inter-amplifier charge transfer efficiency, we proposed and tested several methods to quantify the ability for charge to shift from one amplifier to the next across a dynamic range from $10^3$ to $10^4$ e-. The inter-amplifier charge transfer efficiency was consistent with unity until close to a high signal level of $\sim 50,000$ e-, limited by a gradual deferral of charge through the stages that is cumulative and presents most prominently in the last output stage. We find that the sense node, PS gate, and first horizontal clock phase are most important for ensuring that charge is removed and shifted away from an output stage once it is measured, and these voltages can be optimized further to obtain a greater dynamic range. We find that this inter-amplifier charge transfer efficiency establishes an effective detector full well capacity that is distinct from the full wells defined by typical serial clock phases or pixels in conventional CCDs. Since charge deferral between amplifiers compounds through the stages and is not observed in the first amplifier, the maximum amount of charge that can be efficiently read out through the multiple amplifiers is the effective full well for the MAS CCD. Finally, we measure the nonlinearity of the MAS CCD from 400 e- to nearly 40,000 e- to be $\pm 2.5\%$ in this dynamic range, which is consistent for each amplifier stage and comparable to previous generations of LBNL CCDs.

The progress presented here is the beginning of an extensive effort to fully optimize a MAS CCD as a candidate detector for future spectroscopic experiments and space-based imaging such as HWO. Similar results with respect to noise and inter-amplifier charge transfer efficiency have also been found with a 16 channel MAS CCD operating with LTA readout electronics \citep{10521851,2024arXiv240519505L}. With a fundamental proof-of-concept presented here, upcoming large format, high channel count designs have been pushed forward for prototyping, and new readout electronics solutions that accommodate high channel counts including ASICs are being explored \citep[e.g.,][]{2018SPIE10709E..0TL,10102286}.

\begin{acknowledgments}
    Acknowledgments: This research was supported by the Laboratory Director's Research and Development (LDRD) Program at Lawrence Berkeley National Laboratory under Contract DE-AC02-05CH11231. KL was partially supported by the U.S. Department of Energy, Office of Science, Office of Workforce Development for Teachers and Scientists, Office of Science Graduate Student Research (SCGSR) program. The SCGSR program is administered by the Oak Ridge Institute for Science and Education for the DOE under contract number DE‐SC0014664. ADW was partially supported by Fermilab LDRD (2019.011 and 2022.053), NASA APRA (80NSSC22K1411), and a grant from the Heising-Simons Foundation (\#2023-4611). The multi-amplifier sensing (MAS) CCD was developed as a collaborative endeavor between Lawrence Berkeley National Laboratory and Fermi National Accelerator Laboratory. Funding for the design and fabrication of the MAS device described in this work came from a combination of sources including the DOE Quantum Information Science (QIS) initiative, the DOE Early Career Research Program, and the Laboratory Directed Research and Development Program at Fermi National Accelerator Laboratory under Contract No. DE-AC02-07CH11359. We thank the CCD group at Fermilab for providing the device that was the subject of this study.
\end{acknowledgments}

\software{Astropy \citep{2013A&A...558A..33A,2018AJ....156..123A}, 
            SciPy \citep{2020SciPy-NMeth}, 
          Source Extractor \citep{1996A&AS..117..393B,Barbary2016}
          }

\vspace{5mm}

\bibliography{lbnlmas}{}
\bibliographystyle{aasjournal}

\end{document}